\documentclass[preprint,12pt]{elsarticle}

\usepackage{amssymb}
\usepackage{url}

\journal{NIM-A}

\begin{document}

\begin{frontmatter}



\title{Fabrication of $^{108}$Cd target for the astrophysical p-process studies}


\author[a,b]{Sukhendu Saha}
\author[a,b]{Mousri Paul}
\author[a,b]{Lalit Kumar Sahoo}
\author[a,b]{Dipali Basak}
\author[a,b]{Tanmoy Bar}
\author[c]{Jagannath Datta}
\author[c]{Sandipan Dasgupta}
\author[d]{G.L.N. Reddy}
\author[a]{Chinmay Basu}

\address[a]{Nuclear Physics Division, Saha Institute of Nuclear Physics, 1/AF Bidhannagar, Saltlake, Kolkata-700064, India}
\address[b]{Homi Bhabha National Institute, Anushaktinagar, Mumbai-400094, India}
\address[c]{Analytical Chemistry Division, Bhabha Atomic Research Centre, Variable Energy Cyclotron Centre, 1/AF Bidhannagar, Saltlake, Kolkata-700064, India}
\address[d]{National Centre for Compositional Characterization of Materials, BARC, ECIL Post, Hyderabad-500062, India}

\begin{abstract}
The detailed process of preparing enriched $^{108}$Cd targets on mylar and copper backing using the vacuum evaporation technique is described. These targets were employed in an experiment to measure the proton capture cross-section at energies significantly below the Coulomb barrier, for the astrophysical p-process studies \cite{rauscher2013constraining}\cite{1-gyurky2006106}. Due to the low melting point and high vapor pressure of cadmium, some adjustments were implemented in the Telemark multipocket e-beam setup. The target thickness was determined through the measurement of alpha particle energy loss from a triple alpha source and also by RBS measurements. The thickness of the $^{108}$Cd films varies between 290 to 660 $\mu$g/cm$^2$, with a non-uniformity of approximately 10\%. X-ray Photoelectron Spectroscopy (XPS) and X-ray Fluorescence (XRF) analyses were conducted to examine the presence of impurities and to assess surface morphology, phase, and chemical composition.

\end{abstract}




\end{frontmatter}


\section{Introduction}
\label{}
In essence, a laboratory-based nuclear reaction is characterized by the acceleration of mono-energetic particles, referred to as projectiles, which are directed towards a target system consisting of other elements, which can take the form of foil, pellet, or gas system. The resulting products from this interaction are then detected using specialized detectors. 

Effective target preparation plays a pivotal role in the success of nuclear reaction cross-section measurements, with critical considerations of purity, composition, thickness, and uniformity. In the context of astrophysical reactions, where reaction cross-sections are of the order of nano-barns to pico-barns, the use of thin and enriched targets becomes essential for accurate cross-section measurements \cite{broggini2010luna}\cite{iliadis2015nuclear}.

Nuclear reaction measurements can be categorized as either online or offline, depending on the resulting product nuclei and the specific measurement objectives. Online experiments, requiring charge particle measurements, necessitate a thin target with a thickness of approximately 10-100 $\mu$g/cm$^2$. For neutron or gamma measurements, thicker targets of several hundreds of $\mu$g/cm$^2$ to mg/cm$^2$ can be employed \cite{stolarz2014target}\cite{leo2012techniques}. If the final product is radioactive with a sufficiently long half-life, typically ranging from a few minutes to days, an offline experiment can be conducted to measure the total reaction cross-section. In the case of offline measurements, it is essential for the product nuclei to remain within the target following the projectile bombardment \cite{gyurky2019activation}. Using a target with a catcher foil is a preferable option for these reactions as it helps minimize the loss of product nuclei.

Various methods can be employed for target preparation, including vacuum evaporation and condensation, electrodeposition, rolling, tablet pressing, and others. The selection of the target preparation technique depends on factors such as the material composition, physical form, desired target thickness, uniformity, as well as the purity and availability of the target material \cite{stolarz2014target}.

This paper provides a comprehensive description of the preparation of enriched $^{108}$Cd deposited on a mylar (H$_8$C$_{10}$O$_4$) foil for the measurement of $^{108}$Cd(p,$\gamma$)$^{109}$In cross-section, using the activation technique \cite{gyurky2019activation}. Additionally, another target consisting of enriched $^{108}$Cd on a copper backing was prepared for the investigation of proton elastic scattering near the Coulomb barrier. Both targets were fabricated using vacuum evaporation and condensation method.

The purpose of preparing $^{108}$Cd targets was to study Proton capture reaction $^{108}$Cd(p,$\gamma$) using offline activation technique. The inverse reaction occurs in supernova known as the $\gamma$-process and the experimental determination of the cross-section of this reaction carries great importance in astrophysics \cite{rauscher2013constraining}\cite{1-gyurky2006106}\cite{gyurky2006alpha}.

\section{Deposition setup}
\label{}
Two distinct types of targets were prepared: with depositing Cd on mylar backing, and the other on copper backing. The target fabrication process consists of two primary steps: preparing the backing material and depositing cadmium onto it. For the copper backing, a self-supporting copper layer was deposited using e-beam evaporation and condensation techniques. A commercial mylar foil, with a thickness of 14.3 $\mu$m, was used for the mylar backing.

\begin{figure}[!h] 
	\centering 
	\begin{minipage}[t]{5cm} 
		\centering 
		\includegraphics[scale=0.355]{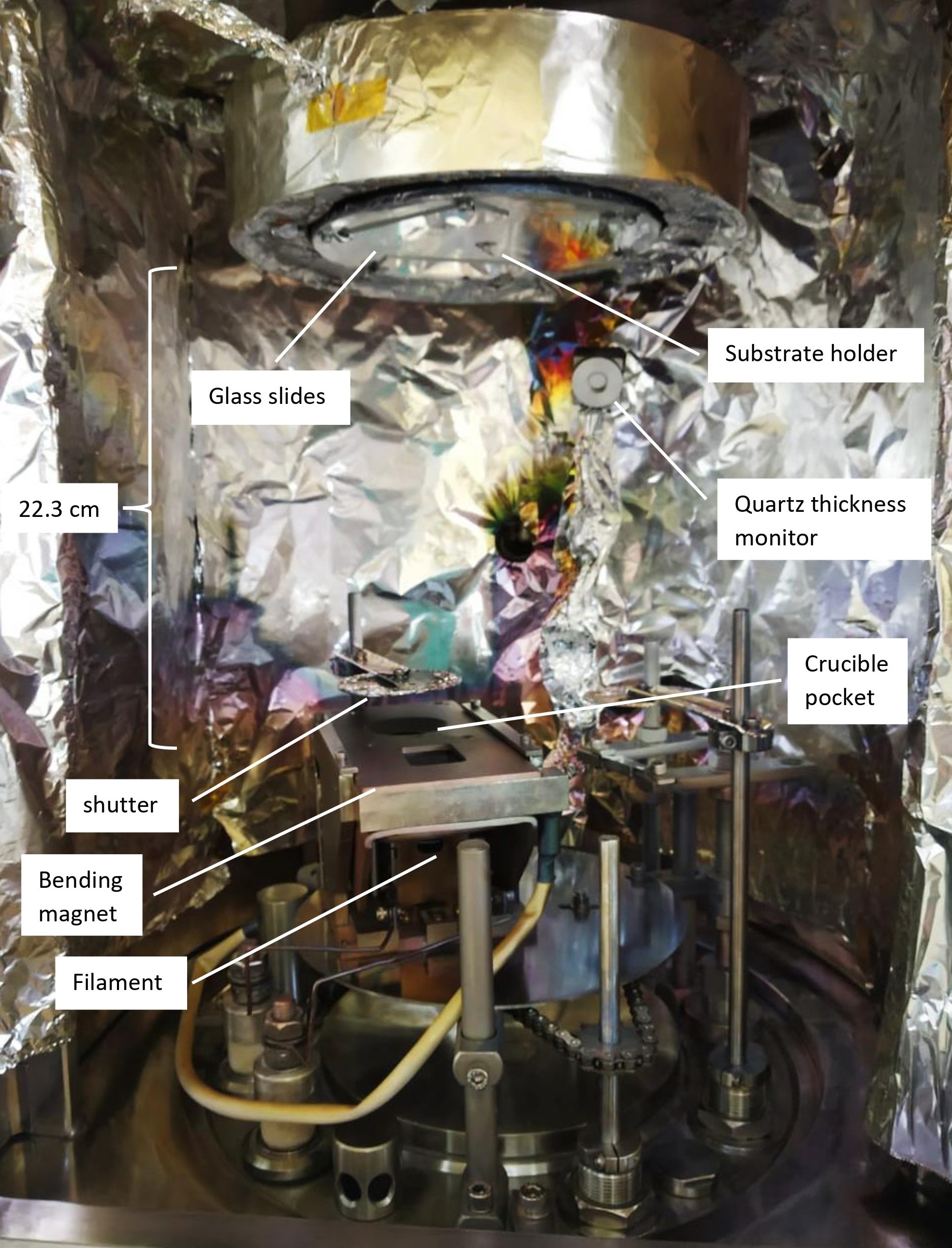} 
		\caption{HHV's Smart Coat 3.0A setup}
        \label{fig1}
	\end{minipage} 
	\hspace{1cm} 
	\begin{minipage}[t]{5cm} 
		\centering 
		\includegraphics[scale=0.35]{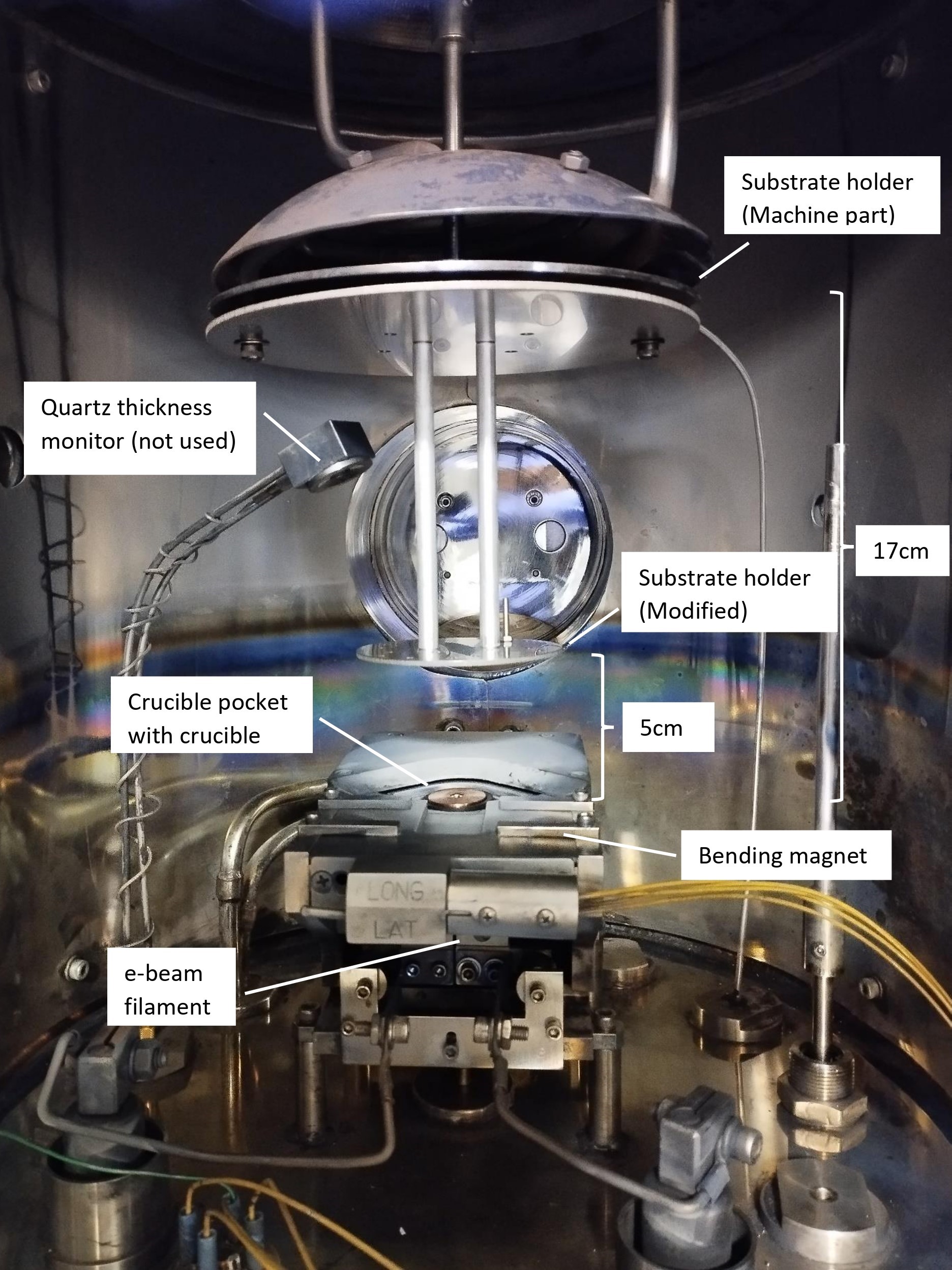} 
		\caption{Telemark multipocket e-beam setup} 
        \label{fig2}
	\end{minipage} 
 
	\begin{minipage}[t]{5cm} 
		\centering 
		\includegraphics[scale=0.55]{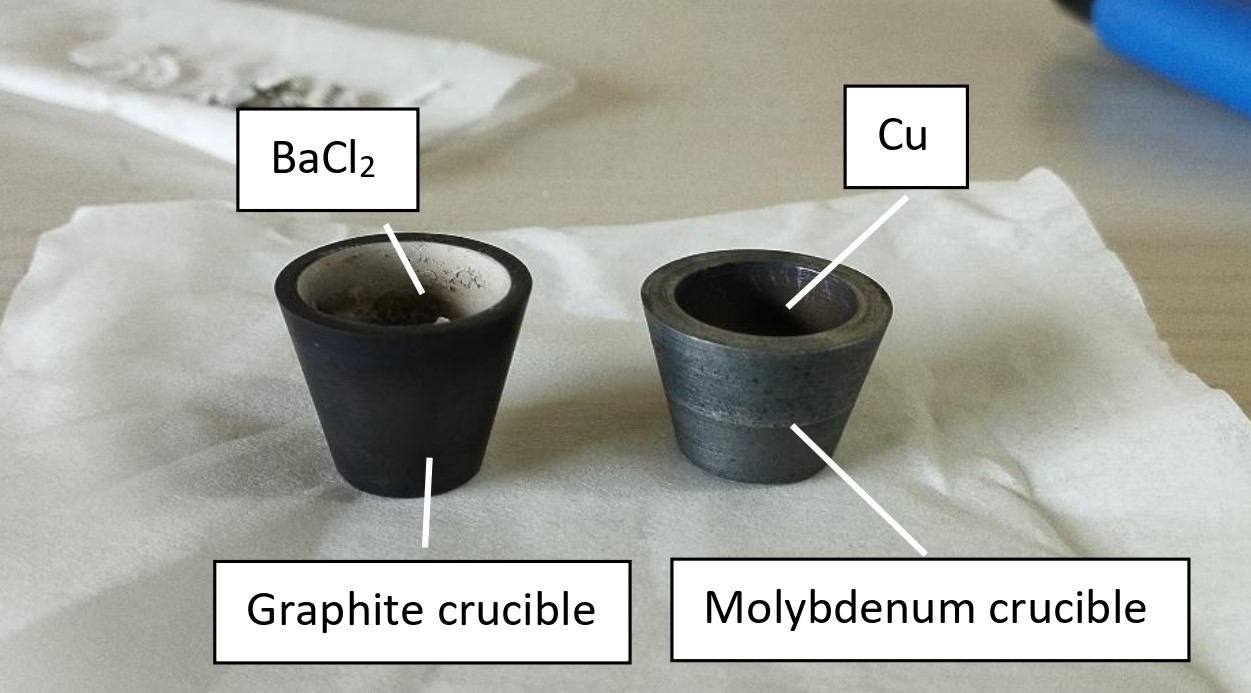} 
		\caption{5cc crucibles has been used for the deposition od BaCl$_2$ and Cu} 
        \label{fig3}
	\end{minipage} 
	\hspace{1cm} 
	\begin{minipage}[t]{5cm} 
		\centering 
		\includegraphics[scale=0.55]{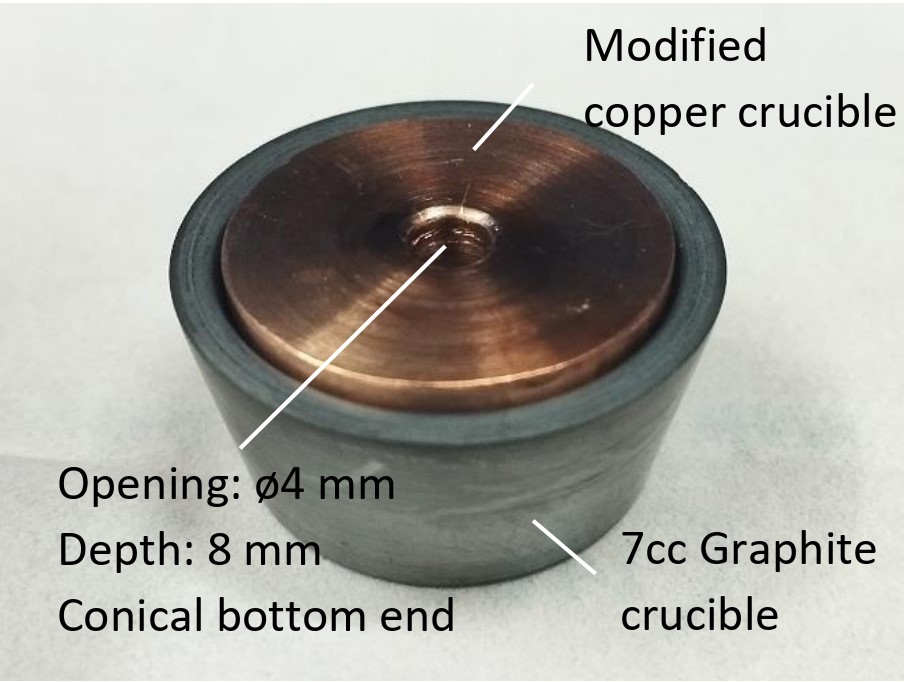} 
		\caption{Modified Cu crucible used for Cd deposition} 
        \label{fig4}
	\end{minipage} 
\end{figure}

The deposition of self-supporting copper was performed using the `Hind High Vacuum Pvt Ltd(HHV) Smart Coat 3.0A' machine as shown in Figure \ref{fig1}. This machine is equipped with three 5cc crucible pockets, and for copper evaporation, a molybdenum crucible was employed, shown in Figure \ref{fig3}. It features a substrate holder positioned 22.3 cm above the crucible pocket, which is attached to a rotating disk. To monitor the thickness of the deposition, a quartz crystal was utilized, and an initial shutter was used to prevent impurities from being deposited onto the substrate. To generate the electron-beam, a tungsten filament was used, and a magnet setup guided the trajectory of the electron-beam through a 270$^\circ$ arc to target the sample, which was maintained at electrical ground potential. The deposited chamber was evacuated using dry roughing (backing) and turbo molecular pump.

\begin{figure}[!h]
     \centering
    \includegraphics[scale=0.45]{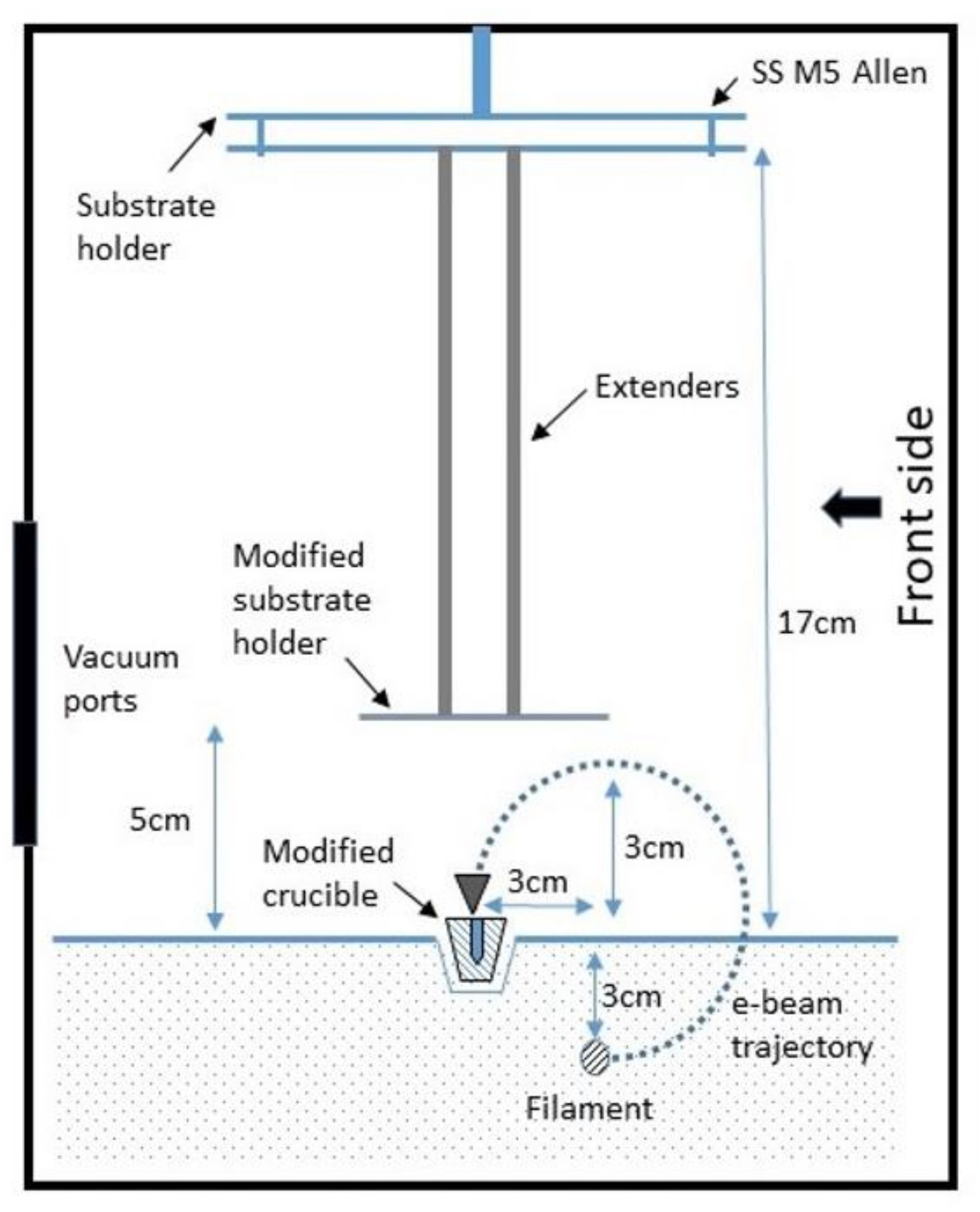}
    \caption{Schematic view of Telemark multipocket e-beam setup}
    \label{fig5}
\end{figure}

For the deposition of cadmium onto backing materials, the `Telemark multipocket e-beam setup' was employed, shown in Figure \ref{fig2}. A graphite crucible compatible with the machine, initially having an opening diameter of $\phi$28 mm, was modified to $\phi$4 mm and a depth of 8 mm, using pure copper, as shown in Figure \ref{fig4}. A pinhole-type modification was also made to reduce the solid angle, resulting in a high-density vapor stream of cadmium. Furthermore, the substrate holder underwent modification to reduce the separation distance from 17 cm to 5 cm from the crucible pocket and was connected to a rotating disk. Notably, the quartz thickness monitor was not utilized due to the reduced separation between the substrate holder and the crucible pocket. Figure \ref{fig5} shows a schematic view of the setup.

\section{Target fabrication methods}
\label{}
Cadmium targets on self-supporting Cu foils and Cd targets on Mylar foils were fabricated using a metal ingot containing 66.3\% enriched $^{108}$Cd. The isotopic distribution provided by the manufacturer has been tabulated in Table \ref{tab1}.

\begin{table}[h!]
    \centering
    \caption{Isotopic distribution of 66.30\% enriched $^{108}$Cd}
    \label{tab1}
    \begin{tabular}{|c|c|c|c|c|c|}
    \hline
       Isotope  & $^{106}$Cd & $^{108}$Cd& $^{110}$Cd& $^{111}$Cd & $^{112,113,114,116}$Cd\\ \hline
       Content(\%)  & 4.6& 66.3 & 29.1 & $<$ 0.0007 & $<$ 0.01\\ \hline
    \end{tabular}
\end{table}

Cadmium does not readily settle onto the substrate due to its high vapour pressure \cite{stolarz2014target}\cite{maier1989special} and low melting point of 321$^\circ$C. As a result, some changes were made to the Telemark multipocket e-beam setup.

\subsection{Deposition of Copper}
The deposition of self-supporting copper foil for the target backing was conducted using the Hind High Vacuum Pvt Ltd Smartcoat 3.0A setup, shown in Figure \ref{fig1}. In this process, three glass slides, each measuring 76 mm$\times$25 mm, were meticulously cleaned with ethanol and were used as substrate.

Before depositing the copper or any self-supporting foil, it is necessary to apply a suitable releasing agent. In this process, BaCl$_2$ was utilized as a water-soluble releasing agent. Initially, BaCl$_2$ pellets were prepared using a hydraulic press. These BaCl$_2$ pallets were then positioned in one graphite crucible, placed in `pocket 1'. Simultaneously, a separate molybdenum (Mo) crucible was used to contain a pure copper metal ingot with a purity level of 99.99\%. This copper filled crucible was placed in `pocket 2'.

The vacuum within the chamber was about 3$\times$10$^{-6}$ mbar. Once the desired pressure was attained, the substrate began to rotate at a speed of 5 revolutions per minute (rpm). Subsequently, a layer of BaCl$_2$ was deposited, with a thickness of approximately $\sim$3 $\mu$m, using e-beam evaporation. Following a one-hour cooling period of the chamber, `pocket 2' was positioned at the e-beam gun point for the copper evaporation. The filament current was gradually increased until a specific deposition rate was initiated. Initially, the deposition rate was 0.1 $\AA$/s, when the e-beam current was 60 mA and then slowly increased to 80 mA in 2 mA increments. At this point, the deposition rate reached 2.5 $\AA$/s. The evaporation process continued until a certain copper thickness, approximately 0.2 $\mu$m, was achieved, as monitored by a quartz crystal. Subsequently, the chamber was allowed to cool before venting.
\begin{table}[h!]
    \centering
    \caption{Melting points, current required and the setup used for the e-beam evaporation of different materials}
    \label{tab2}
    \begin{tabular}{|c|c|c|c|}
    \hline 
    & MP($^oC$)& Setup used &Current(mA) \\ \hline
    BaCl$_2$& 962 & HHV's Smart Coat 3.0A  &10-15 \\
    Cd& 321.1 & Telemark e-beam setup &3-7 \\
    Cu& 1085 & HHV's Smart Coat 3.0A &60-80 \\
    \hline
    \end{tabular}
\end{table}

\subsection{Fishing and mounting the backing on target frame}

\begin{figure}[!h] 
	\centering 
	\begin{minipage}[t]{5cm} 
		\centering 
		\includegraphics[scale=0.43]{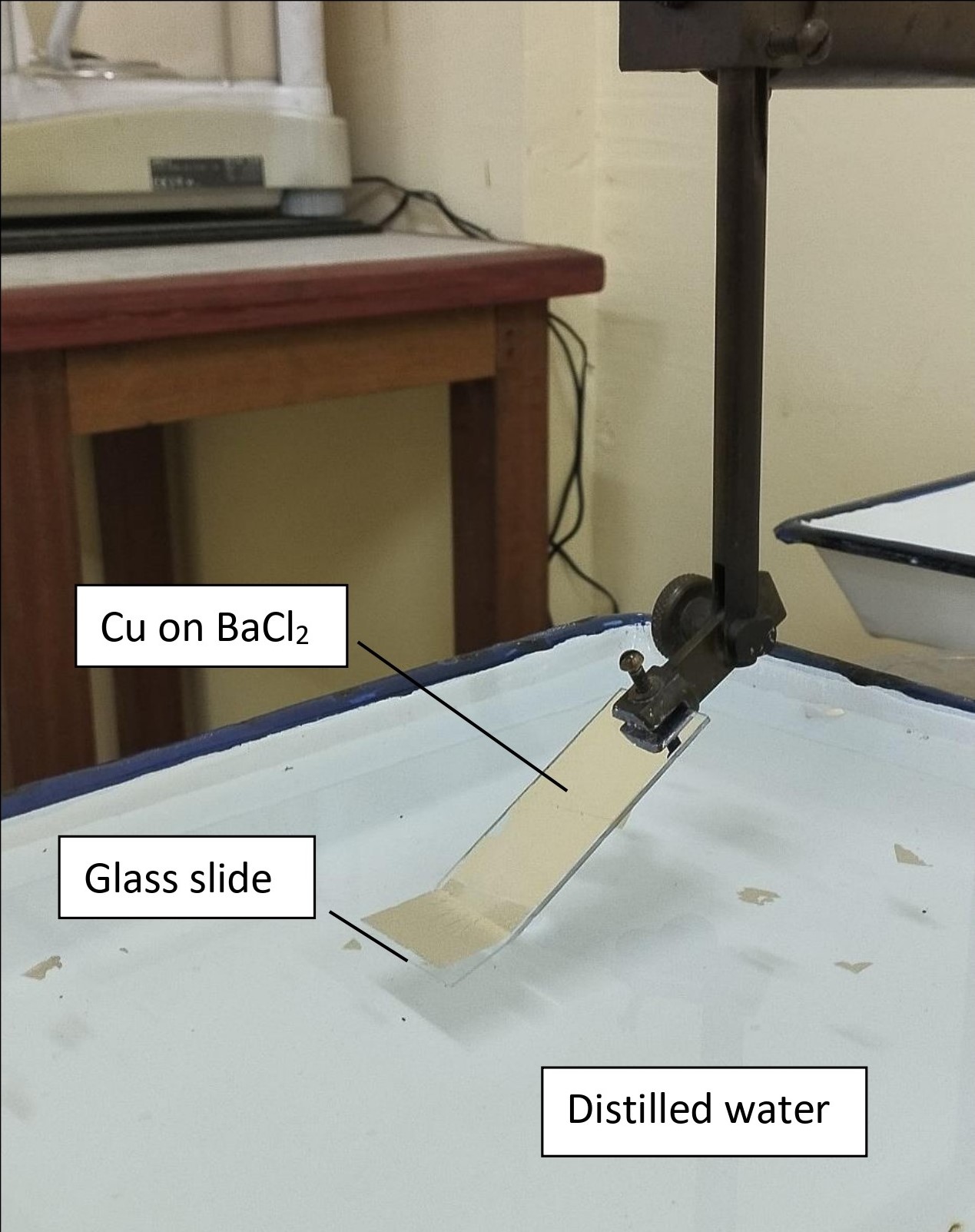} 
		\caption{Floating of Cu foils}
        \label{fig6}
	\end{minipage} 
	\hspace{1cm} 
	\begin{minipage}[t]{5cm} 
		\centering 
		\includegraphics[scale=0.4]{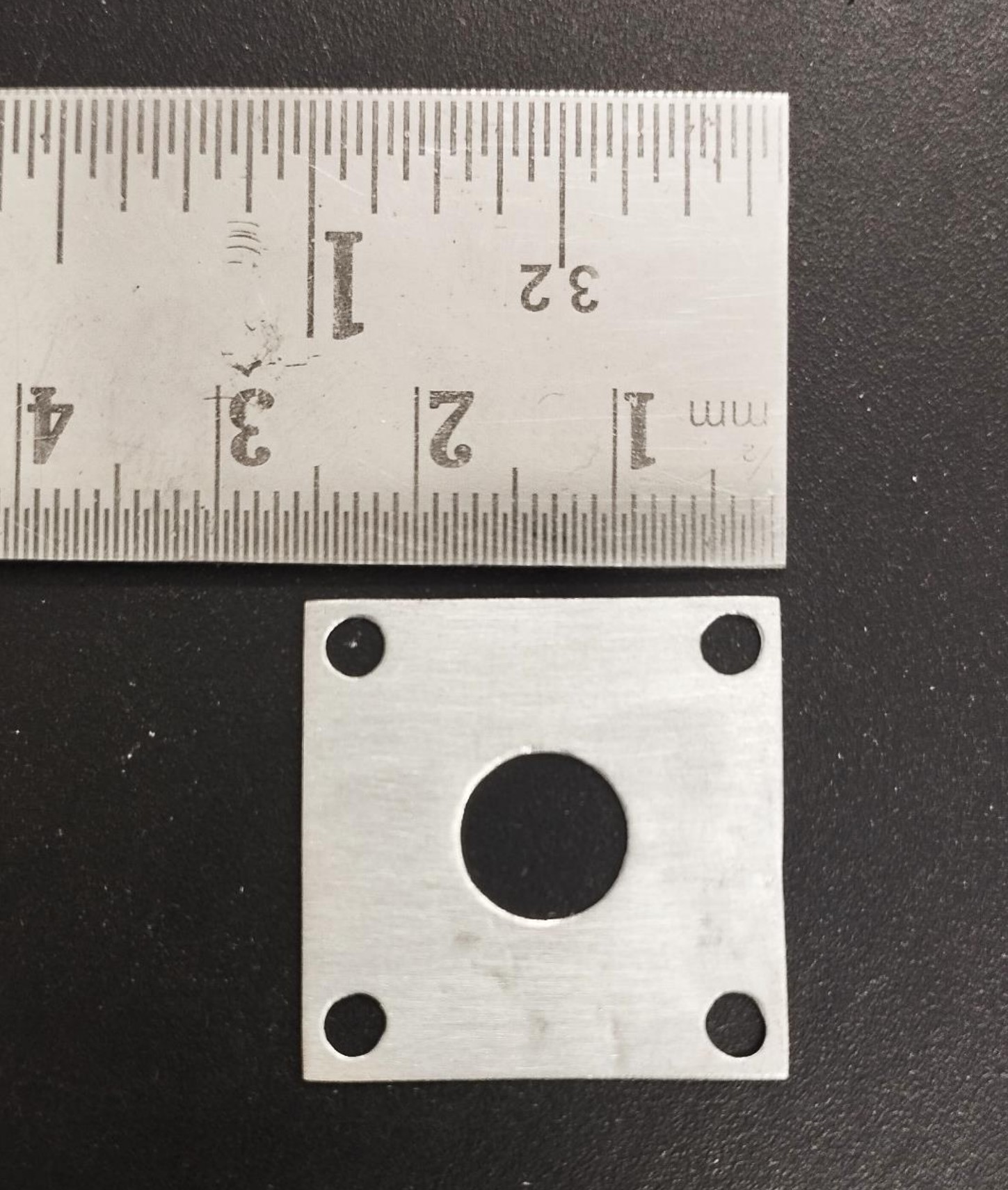} 
		\caption{Target frames used to catch the Cu foils} 
        \label{fig7}
	\end{minipage} 
 \end{figure}
The glass slides were carefully removed and gently immersed in a distilled water bath at an angle of $\sim$35$^\circ$ with respect to water surface, as shown in Figure \ref{fig6}. When BaCl$_2$ dissolved in water, the copper foil floated on the water's surface. Target frames, each with dimensions of 26 mm$\times$26 mm and a central hole of diameter approximately $\phi$10 mm, as illustrated in Figure \ref{fig7}, were used to collect the Cu foil. After allowing sufficient time for the foils to dry completely, the thickness of the Cu foil was measured using the energy loss of a known triple-alpha source, detected with 60 $\mu$m Si surface barrier detectors. The details of the thickness measurement are discussed in Section \ref{thickness}.

\subsection{Heat test for Mylar backing}
\begin{figure}[!h] 
	\centering 
	\begin{minipage}[t]{5cm} 
		\centering 
		\includegraphics[scale=0.45]{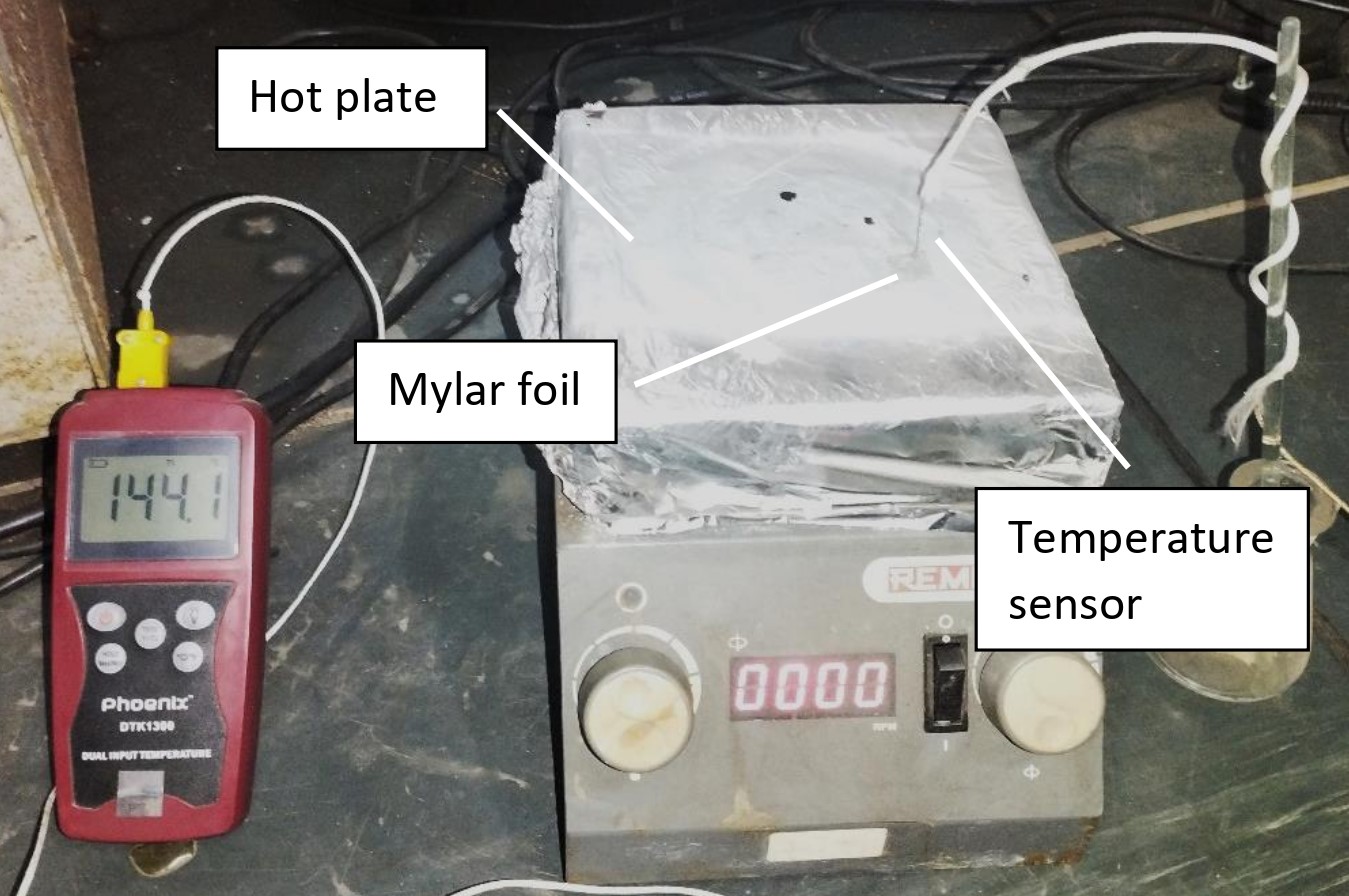} 
		\caption{Heat test of Mylar foils using hotplate}
        \label{fig8}
	\end{minipage} 
	\hspace{1cm} 
	\begin{minipage}[t]{5cm} 
		\centering 
		\includegraphics[scale=0.45]{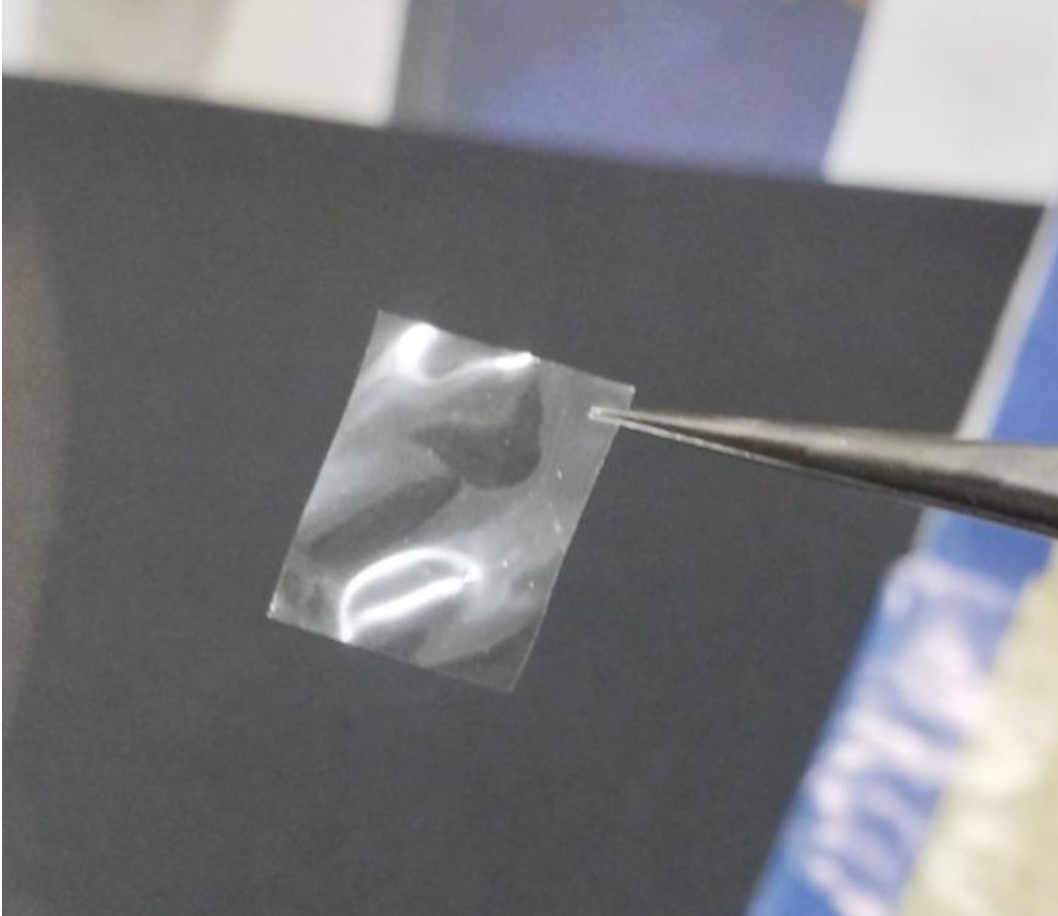} 
		\caption{Deformation of Mylar foil at 155$^\circ$C} 
        \label{fig9}
	\end{minipage} 
 \end{figure}
While the melting point of Cd is 321 $^\circ$C, Mylar foil has a lower melting point of 254 $^\circ$C. Mylar foil can maintain its shape up to approximately 150 $^\circ$C. however, it starts to deform above this temperature, as depicted in Figure \ref{fig8} and \ref{fig9}. To assess its sustainability at high temperatures, heat tests were conducted using both an electric hot plate and an infrared (IR) lamp. An electric thermometer was employed to monitor the temperature. The results indicated that the Mylar foil retained its shape for an extended period at temperatures up to 150 $^\circ$C. Beyond this temperature, the foil exhibited deformation.

\subsection{Deposition of Cd on Cu and Mylar backing}
The deposition of Cd on the backing material was carried out using the Telemark multipocket e-beam setup. For this process, a 38.7 mg sample of enriched $^{108}$Cd metal was carefully placed in the modified Cu crucible, as shown in Figure \ref{fig4}. Prior to the deposition, the crucible and the entire chamber were meticulously cleaned using ethanol and isopropanol to ensure a pristine environment. In addition, commercial mylar foil was cut into 25 mm square shapes, and its thickness was verified by measuring the alpha particle energy loss as discussed in Section \ref{thickness}. Similarly, the thickness of the self-supporting Cu foil was measured. The mylar foil and the target frame, which contained the Cu foil, were securely attached to the modified substrate holder. This substrate holder was then affixed to the rotating disk.

\begin{figure}[!h] 
	\centering 
	\begin{minipage}[t]{5cm} 
		\centering 
		\includegraphics[scale=0.45]{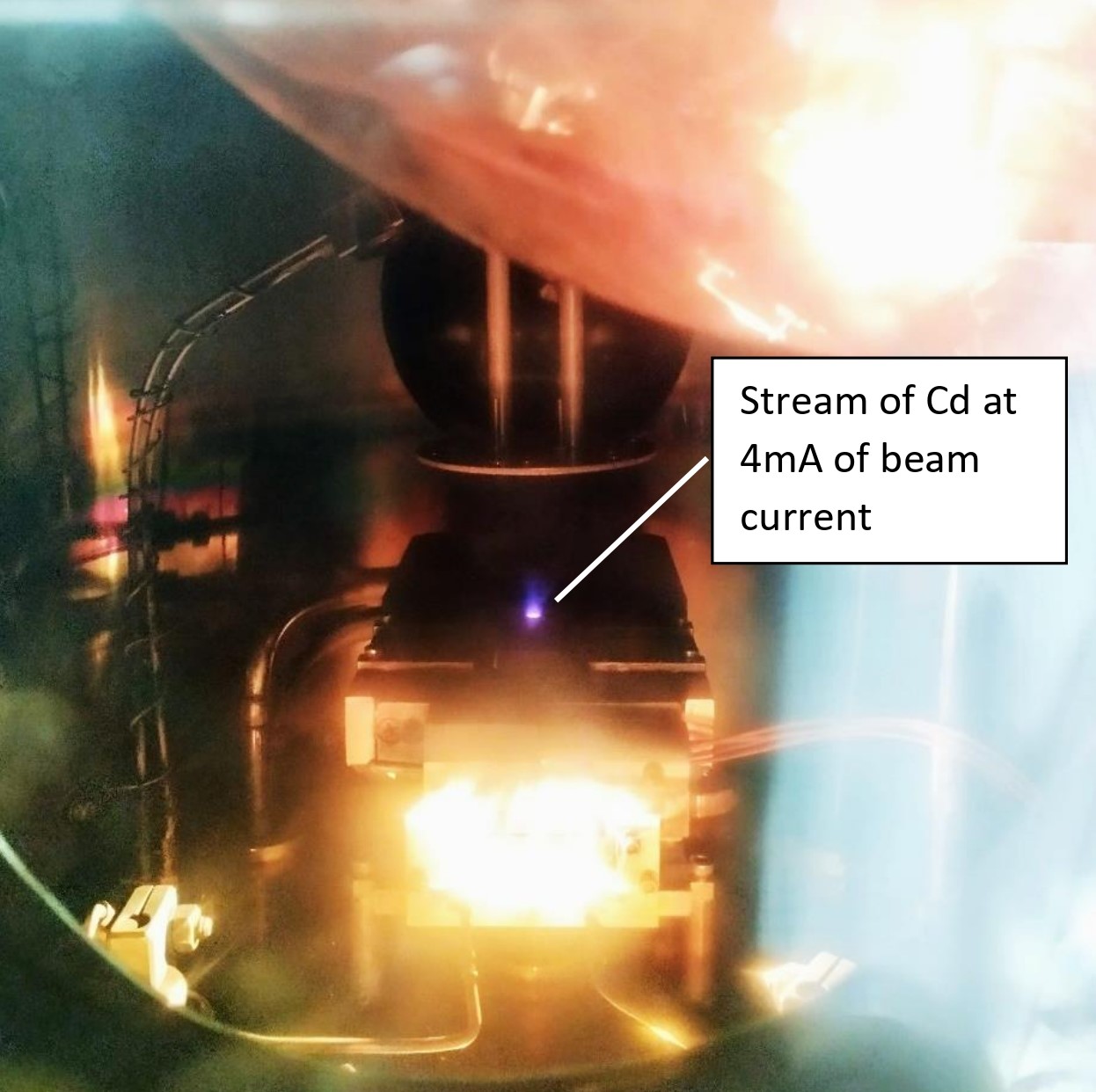} 
		\caption{While evaporation of Cd using Telemark multipocket e-beam setup}
        \label{fig10}
	\end{minipage} 
	\hspace{1cm} 
	\begin{minipage}[t]{5cm} 
		\centering 
		\includegraphics[scale=0.65]{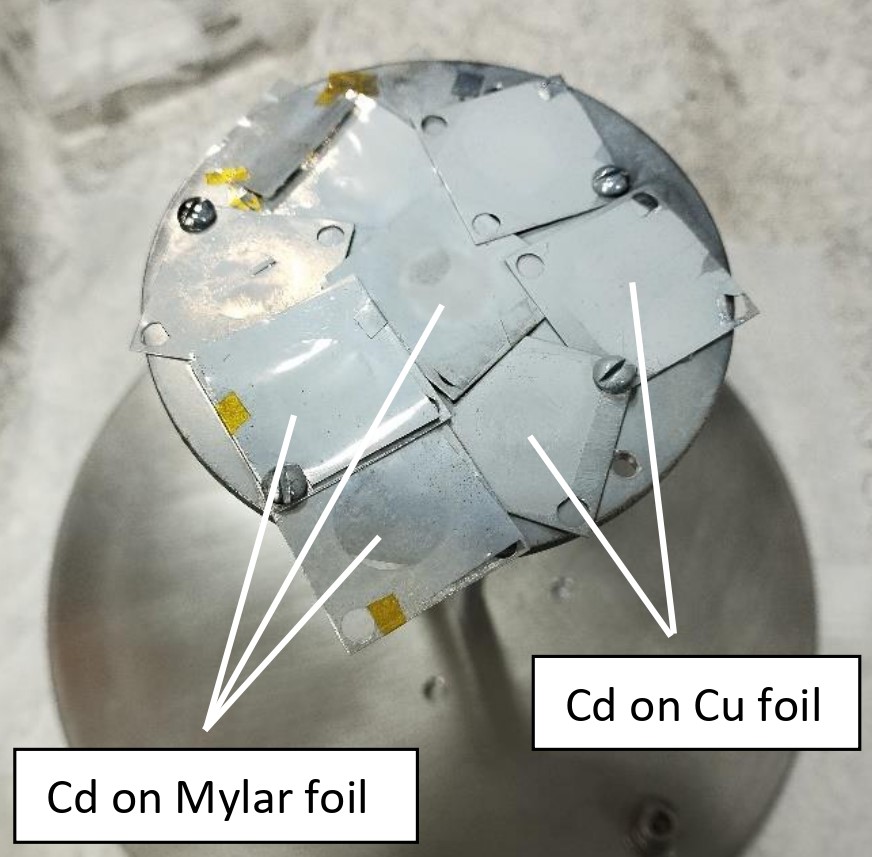} 
		\caption{Deposited Cd onto the backing foil attached to the substrate holder} 
        \label{fig11}
	\end{minipage} 
 \end{figure}

The chamber vacuum 6$\times$10$^{-7}$ mbar. To ensure a uniform Cd deposition, the substrate was set in rotation at 5 rpm. Subsequently, the tungsten filament was powered on with a voltage of 5.95 kV and a current of 17.8 A.  Cd evaporation commenced at a beam current of 3 mA, at that time the chamber pressure reached $\sim$3$\times$10$^{-6}$ mbar, resulting in the visible appearance of a distinctive blue stream, as depicted in Figure \ref{fig10}. The e$^-$-beam current was carefully increased to 10 mA, using 1 mA increments, over a period of 30 minutes, during which the blue stream gradually dissipated.  After the deposition process was complete, the chamber was allowed to cool for 4 hours. Following this, the chamber was vented, and the targets were carefully removed from the substrate, and subsequently stored in a vacuum desiccator.

\begin{figure}[!h] 
	\centering 
	\begin{minipage}[t]{5cm} 
		\centering 
		\includegraphics[scale=0.5]{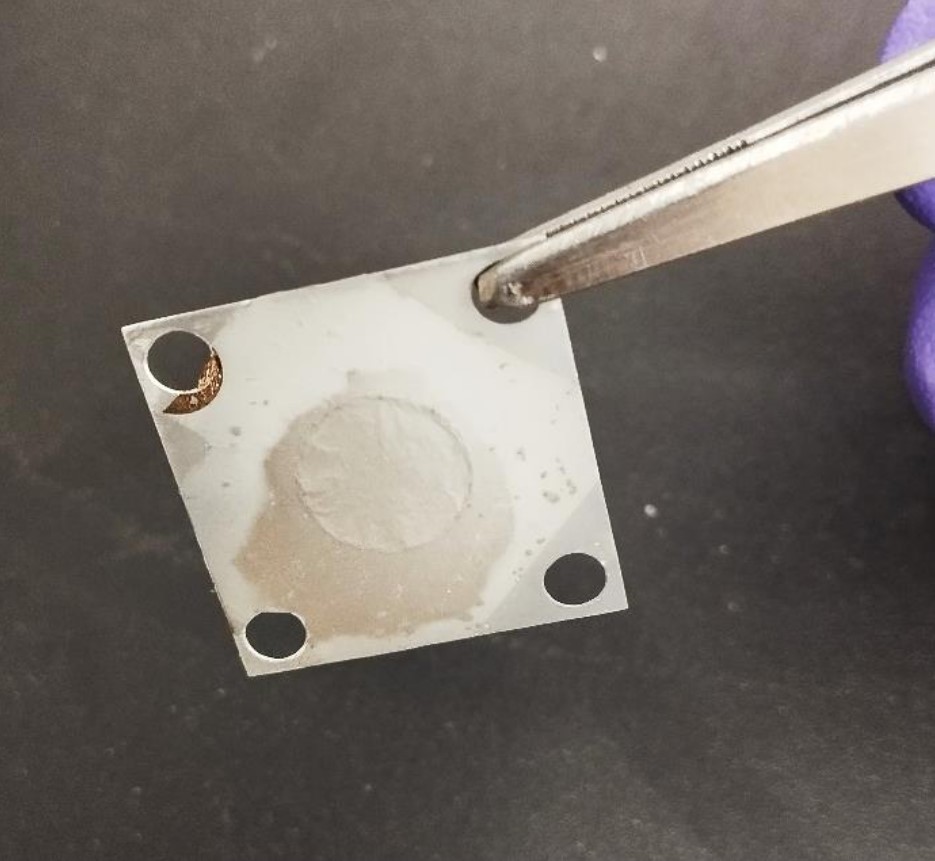} 
	\end{minipage} 
	\hspace{0.5cm} 
	\begin{minipage}[t]{5cm} 
		\centering 
		\includegraphics[scale=0.54]{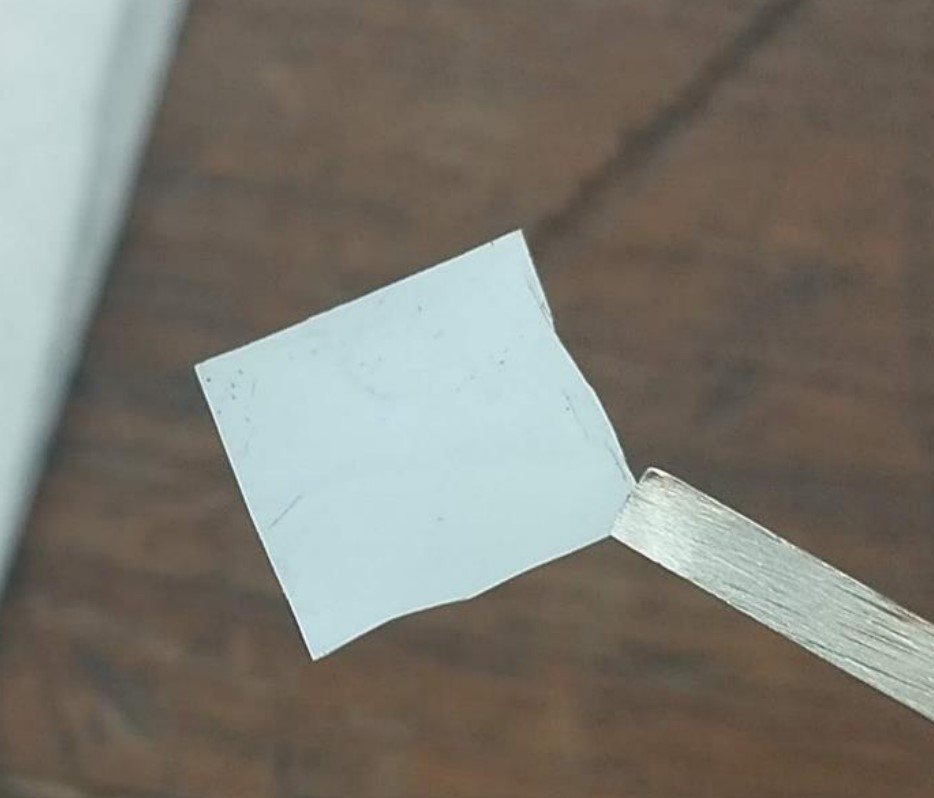} 
	\end{minipage}
 \caption{Cd deposited on Cu backing and Cd deposited on Mylar backing targets}
 \end{figure}
 
Due to the highly toxic nature of cadmium, rigorous safety precautions were meticulously observed throughout the deposition process \cite{guideline}.

\section{Characterisation of the targets}
\subsection{Thickness measurement using triple-$\alpha$ source}
\label{thickness}
\begin{figure}[!h] 
	\centering 
	\begin{minipage}[t]{5cm} 
		\centering 
		\includegraphics[scale=0.45]{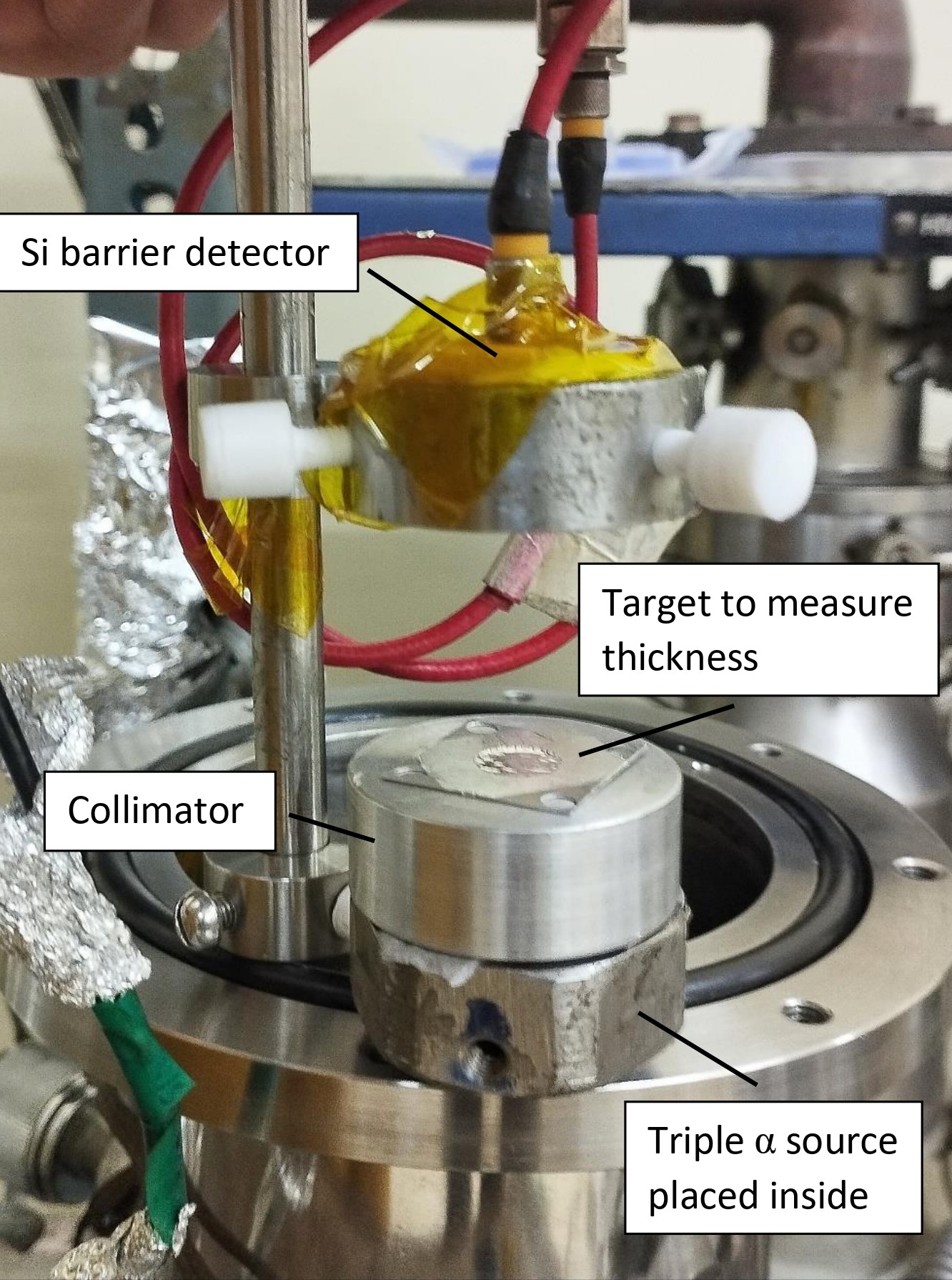} 
		\caption{Thickness measurement setup by detecting alpha energy loss by target foil}
        \label{fig12}
	\end{minipage} 
	\hspace{1cm} 
	\begin{minipage}[t]{5cm} 
		\centering 
		\includegraphics[scale=0.6]{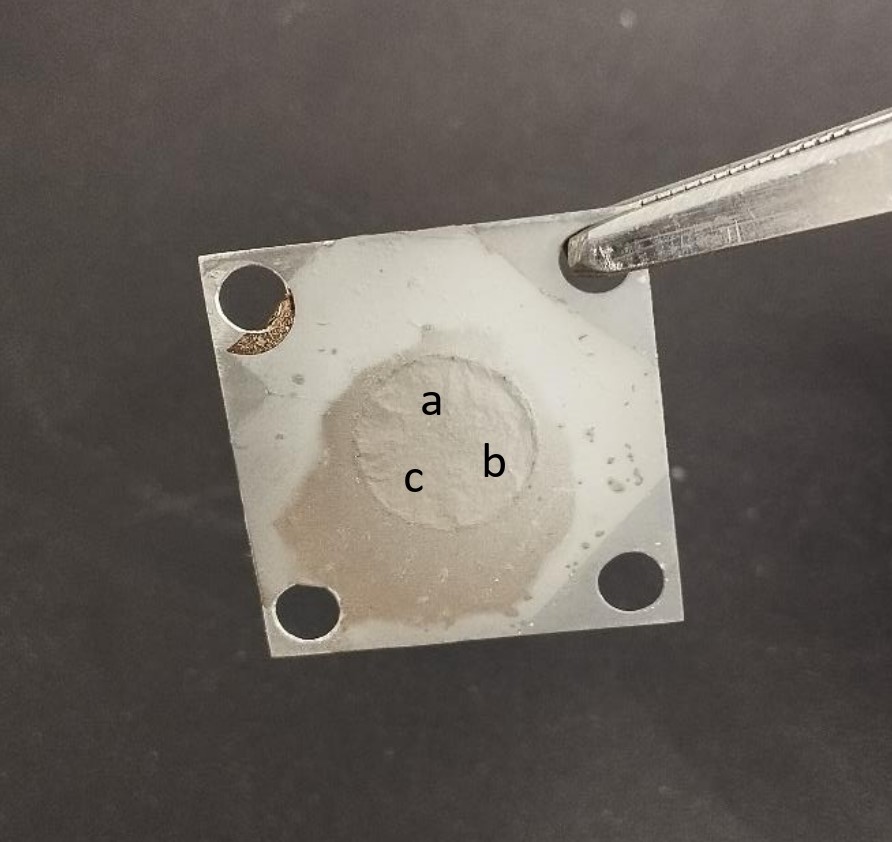} 
		\caption{Thickness measurement at different positions of the target} 
        \label{fig13}
	\end{minipage} 
 \end{figure}

The target thickness was determined in two stages. Initially, the thickness of the target backing was measured prior to the deposition of cadmium. Subsequently, after the Cd deposition on the backing, the final thickness was measured. The thickness of the Cd layer was estimated by subtracting the backing thickness from the final thickness. The experimental setup and schematic diagram are presented in Figure \ref{fig12}.

A triple alpha source (containing $^{239}$Pu, $^{241}$Am and $^{244}$Cm) with three distinct alpha energy lines at 5155 keV, 5486 keV, and 5806 keV \cite{srim-yalcin2015thickness}\cite{sahoo2023preparation}, is positioned just below the collimator. The target foil is positioned above the collimator, allowing for the measurement of a specific section of the target's thickness. A 60$\mu$m Si-surface barrier detector made by EG$\&$G ORTEC, USA was employed to detect alpha particles.

The energies of the three $\alpha$ particles with and without the target were measured using the silicon surface barrier detector. The shift of the energy positions (\textit{$\Delta$E}) is used to determine the thickness (\textit{$\Delta$x}) of the target from the experiment,
\begin{equation}
\Delta x=\frac{\Delta E}{-(dE/dx)}
\end{equation}
where --$\frac{dE}{dx}$ is the stopping power of the target at specific alpha energy, E$_\alpha$. The stopping power is obtained from the code SRIM \cite{srim-ziegler2010srim}.

\begin{figure}[!h] 
	\centering 
	\begin{minipage}[t]{6cm} 
		\centering 
		\includegraphics[scale=0.4]{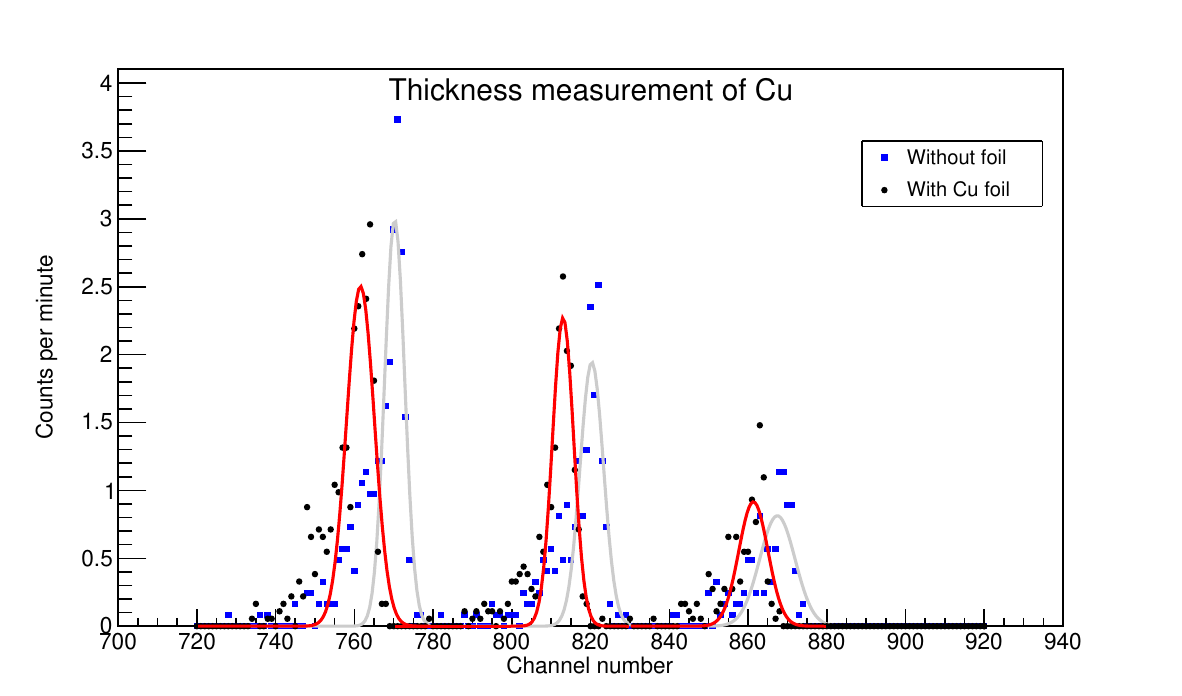} 
        \label{fig14a}
	\end{minipage} 
	\hspace{1cm} 
	\begin{minipage}[t]{6cm} 
		\centering 
		\includegraphics[scale=0.4]{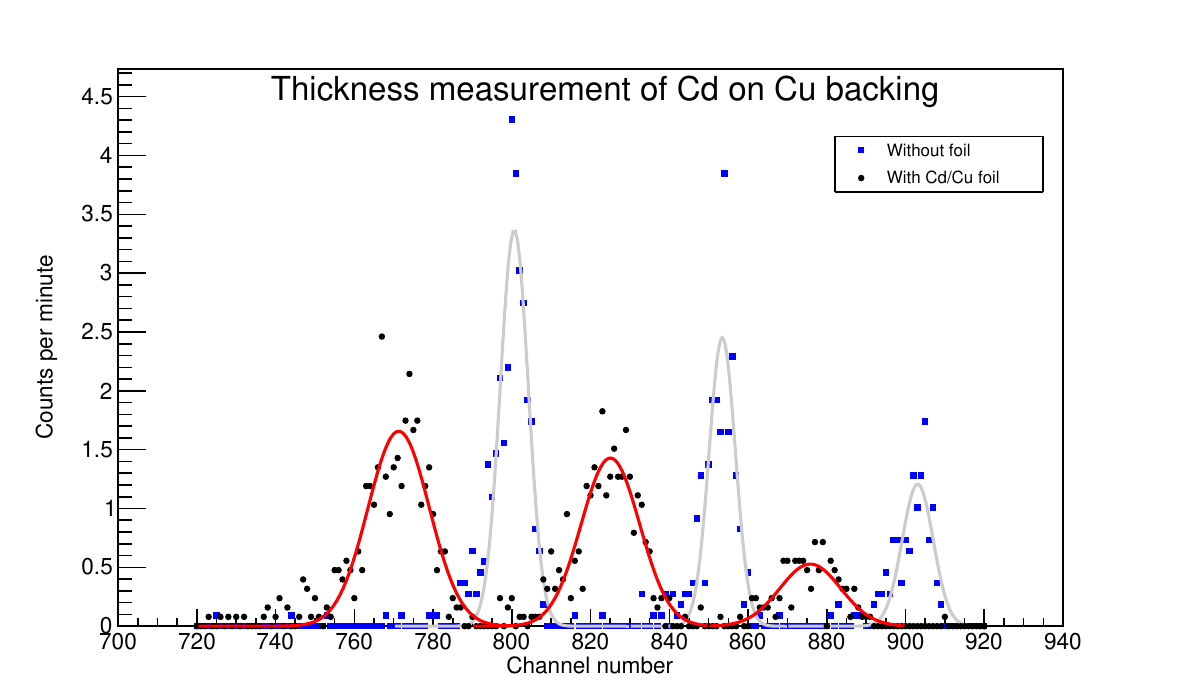} 
        \label{fig14b}
	\end{minipage} 
 \begin{minipage}[t]{6cm} 
		\centering 
		\includegraphics[scale=0.4]{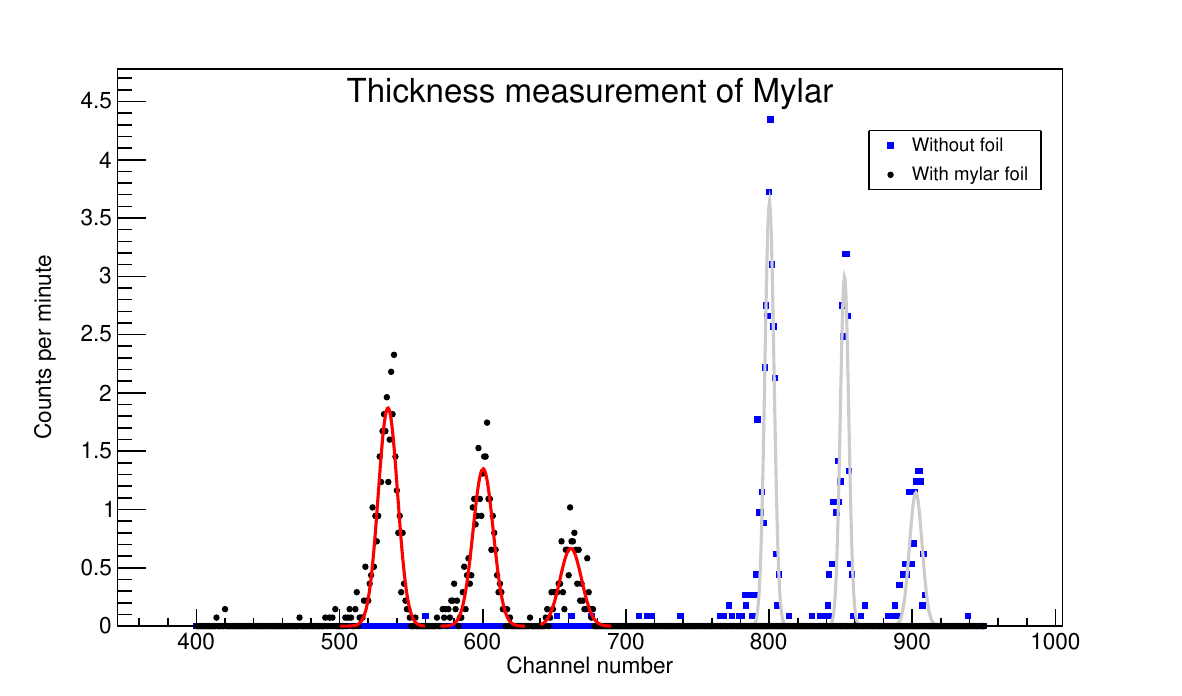} 
        \label{fig14c}
	\end{minipage} 
	\hspace{1cm} 
	\begin{minipage}[t]{6cm} 
		\centering 
		\includegraphics[scale=0.4]{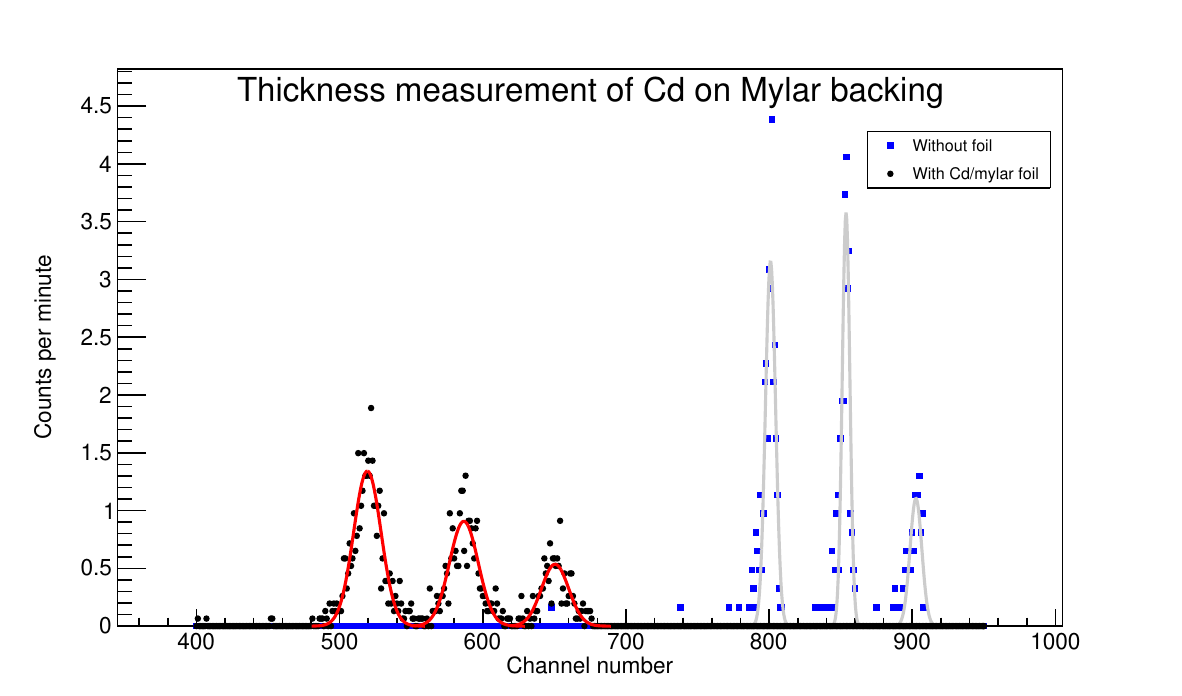} 
        \label{fig14d}
	\end{minipage} 
 \caption{Thickness measurement by measuring energy loss of alpha particle, Cu foil, Cd on Cu foil, Mylar foil and Cd on Mylar foil respectively}
 \end{figure}

The chamber was maintained at a pressure of $\sim$10$^{-5}$ mbar. Data acquisition was performed with an MCA featuring MPANT version 2.1 software, manufactured by FAST ComTec, Germany.

Thickness of the target number `Cd3' (cadmium on mylar) and `317CuCd' (cadmium on copper) has been tabulated in Table \ref{tab3}.
\begin{table}[h!]
    \centering
    \caption{Thickness of Cadmium targets by $\alpha$-energy loss measurement}
    \label{tab3}
    \resizebox{\linewidth}{!}{
    \begin{tabular}{cp{2.2cm}cp{2.6cm}p{2.4cm}p{2.2cm}p{2.2cm}p{2.4cm}p{2.4cm}p{2.2cm}p{2.4cm}l}
    \hline 
    &Backing thickness (material)&$\alpha$ energy(keV)& $\alpha$-energy loss(backing+ Cd) (keV)&$\frac{dE}{dx}$ of backing (keV/$\mu$m)& $\alpha$-energy loss in backing (keV) & $\alpha$-energy loss in Cd (keV)&$\alpha$-energy falling on Cd layer (keV)& $\frac{dE}{dx}$ of Cd (keV/$\mu$m)& Thickness of Cd ($\mu$m)& avg thickness ($\mu$m)\\ \hline
    &&5155&1825.9& 116.0& 1635.6& 190.4 &3519.4&361.6& 0.53&\\
    &14.1$\mu$m (Mylar)&5486 &1728.8& 110.9& 1563.7& 165.2& 3922.3& 343.2& 0.48&0.46&\\
    &&5805&1630.2 & 106.6& 1503.1& 127.1& 4301.2& 327.8& 0.39&\\
    
    \hline
&&5155 & 205.1 & 372.3 & 78.2& 126.9& 5076.8& 300.2& 0.42&\\
&0.21$\mu$m (Cu)& 5486& 200.0 & 359.6& 75.5& 124.5& 5410.5& 290.2& 0.43& 0.43&\\
&&5805& 202.5& 384.6& 80.7& 121.7& 5724.2& 281.5&0.43&&\\
\hline
    
    \end{tabular}
    }
\end{table}

For the uniformity check, the thickness was measured at three different positions on the same target, as shown in Figure \ref{fig13}. These positions, labeled as (a), (b), and (c), allowed the alpha particles to pass through specific 3 mm areas of the target.

\subsection{Rutherford Back-scattering Spectroscopy (RBS)}
The ion beam analysis (IBA) experiments involving backscattering spectrometry (BS)  were carried out using the 3 MV Tandetron (HVE, Europa) at the Surface and Profile Measurement Laboratory, NCCCM, Hyderabad. The thickness of one of the Cd-deposited targets on Mylar was determined using Rutherford Backscattering Spectroscopy (RBS). This involved the use of a 1 MeV proton beam and a 6 MeV Carbon beam, with a beam current of 4 nA.  

\begin{figure}[!h]
     \centering
    \includegraphics[scale=0.8]{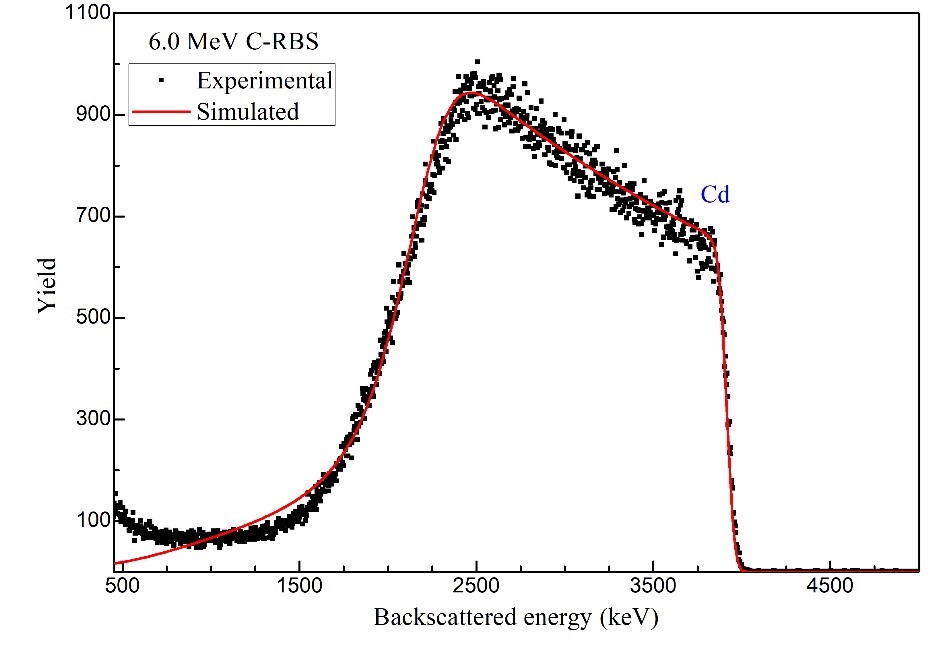}
    \caption{RBS spectrum of enriched $^{108}$Cd target deposited on mylar backing}
    \label{fig17}
\end{figure}

The Backscattering (EBS) measurements involved, bombardment of the samples with a well-collimated beam of proton and carbons  of  energy 1 MeV and 6 MeV, respectively, with a beam current of 5 nA ($\Phi$ = 2.0 mm). The scattered particles were detected at a backward angle of 170$^\circ$ with a Si surface barrier detector. The detector subtends a solid angle of 1.2$\times$10$^{-3}$ sr.  Each spectrum was collected by impinging the sample with about 3.0 $\mu$C of charge, sufficient to produce statistically significant spectra. The spectra were acquired by 8K-PC based MCA.  These were simulated using SIMNRA, a computer code for simulating the energy spectra of charged particles, for qualitative and quantitative analyses \cite{rbs-mayer1997simnra}.

For simulation of RBS spectra using SIMNRA, following conditions were adopted: (a) simulations were continued by refining each layer composition and thickness for the best fit, (b) among the many stopping power data, we have used Ziegler-Biersack data which is more accurate and reliable as the selection of electronic stopping power data has a large influence on the shape of simulated spectra, (c) for Energy-loss straggling, Chu + Yang’s theory was used. The refinement of layer composition was stopped when reduced $\chi^2$ (the quadratic deviation between experimental and simulated data) reached less than 5.

The simulation results show that the thickness of the $^{108}$Cd layer is approximately 487 nm (areal density $\sim$421 $\mu$g/cm$^2$). The measurement come with an associated uncertainty of nearly 10\%.

\subsection{X-ray Photo-electron Spectroscopy (XPS)}

\begin{figure}[!h]  
		\centering 
		\includegraphics[scale=0.3]{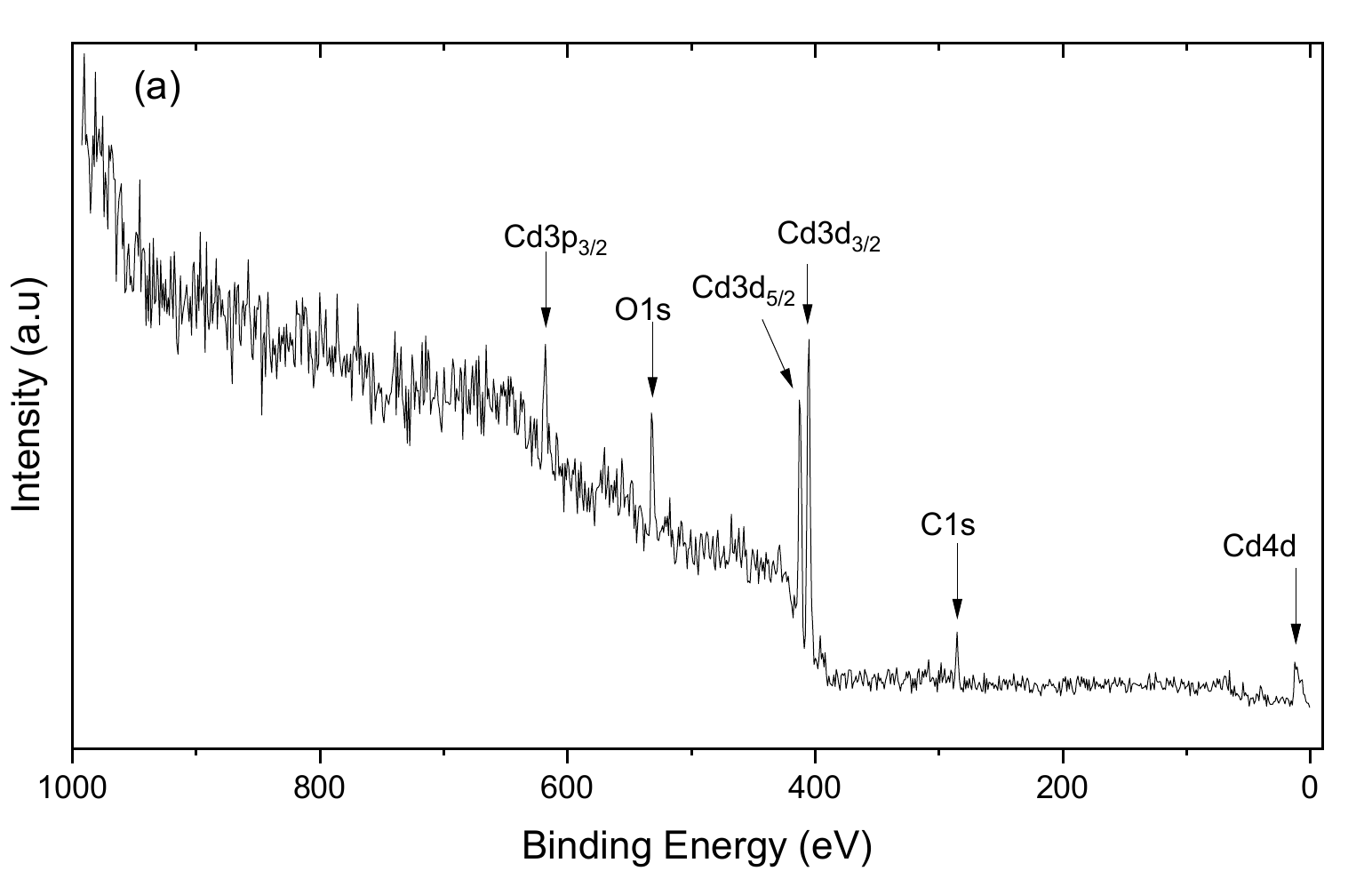} 
        \label{fig15a}
        \caption{(a) Full survey of XPS spectra}
\end{figure}

 \begin{figure}[!h]  
 \centering
	\begin{minipage}[t]{6cm} 
		\centering 
		\includegraphics[scale=0.3]{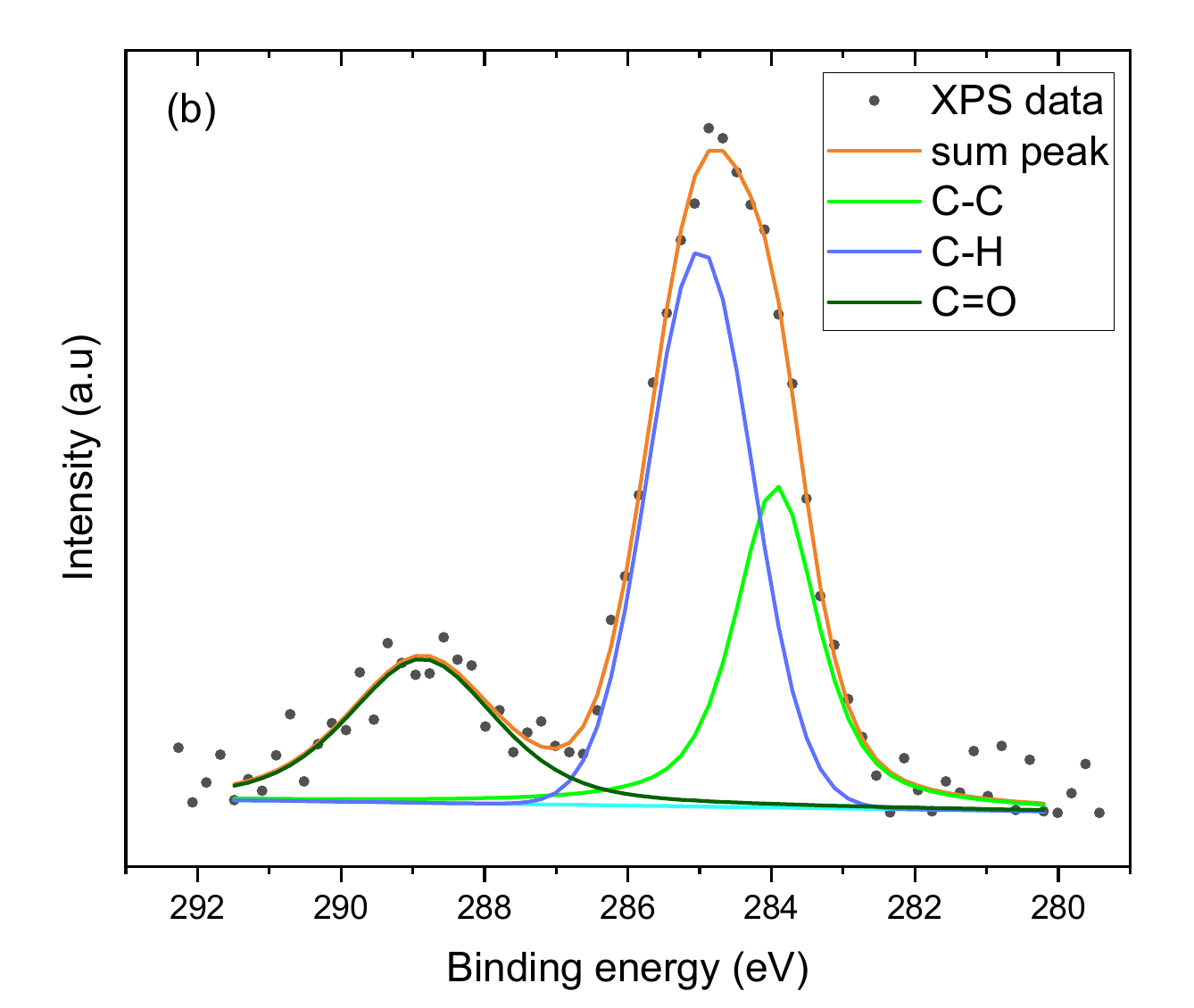} 
        \label{fig15b}
	\end{minipage}
 \hspace{1cm}
        \begin{minipage}[t]{6cm} 
		\centering 
		\includegraphics[scale=0.3]{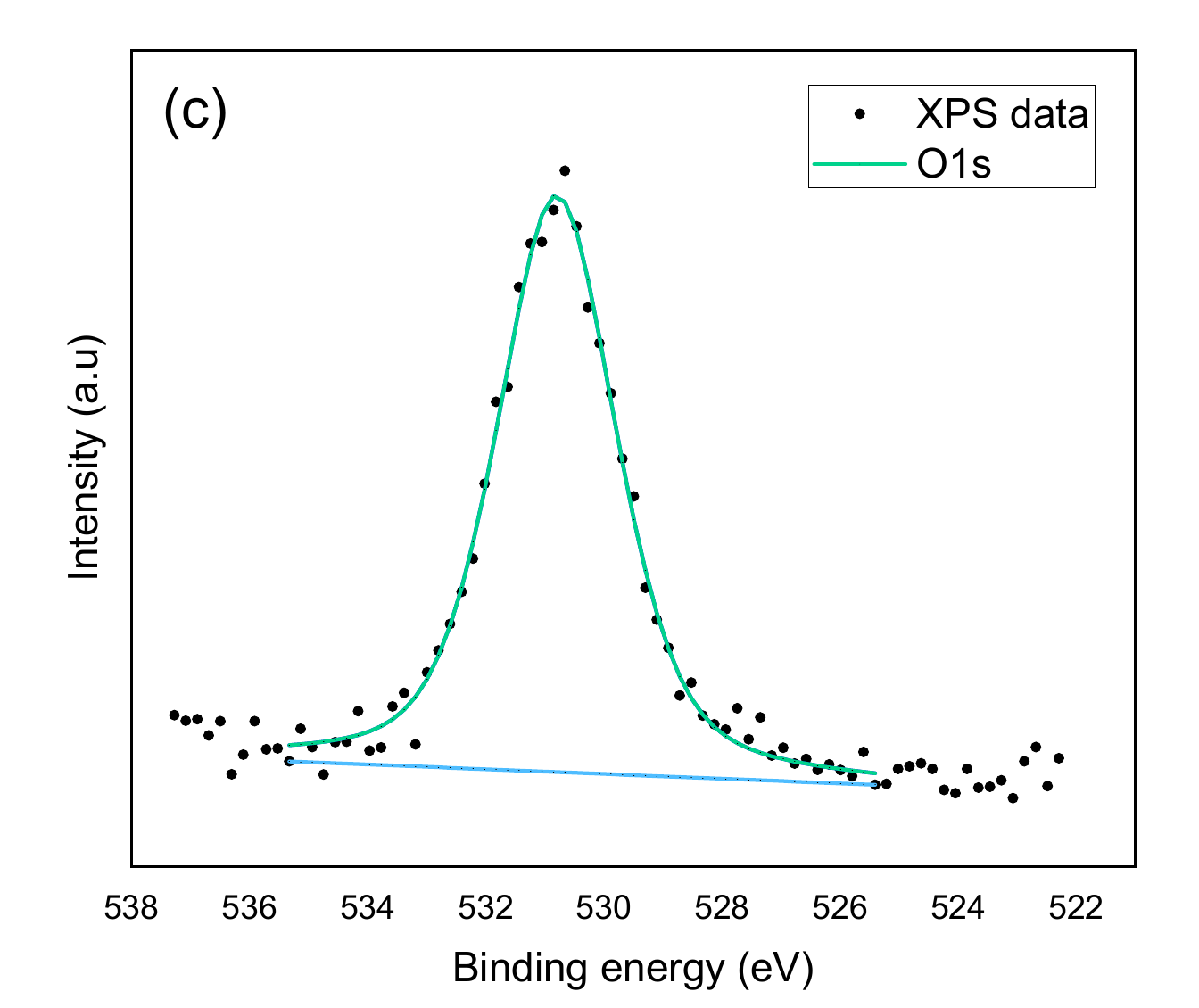} 
        \label{fig15c}
	\end{minipage} 
\hspace{1cm} 
	\begin{minipage}[t]{6cm} 
		\centering 
		\includegraphics[scale=0.3]{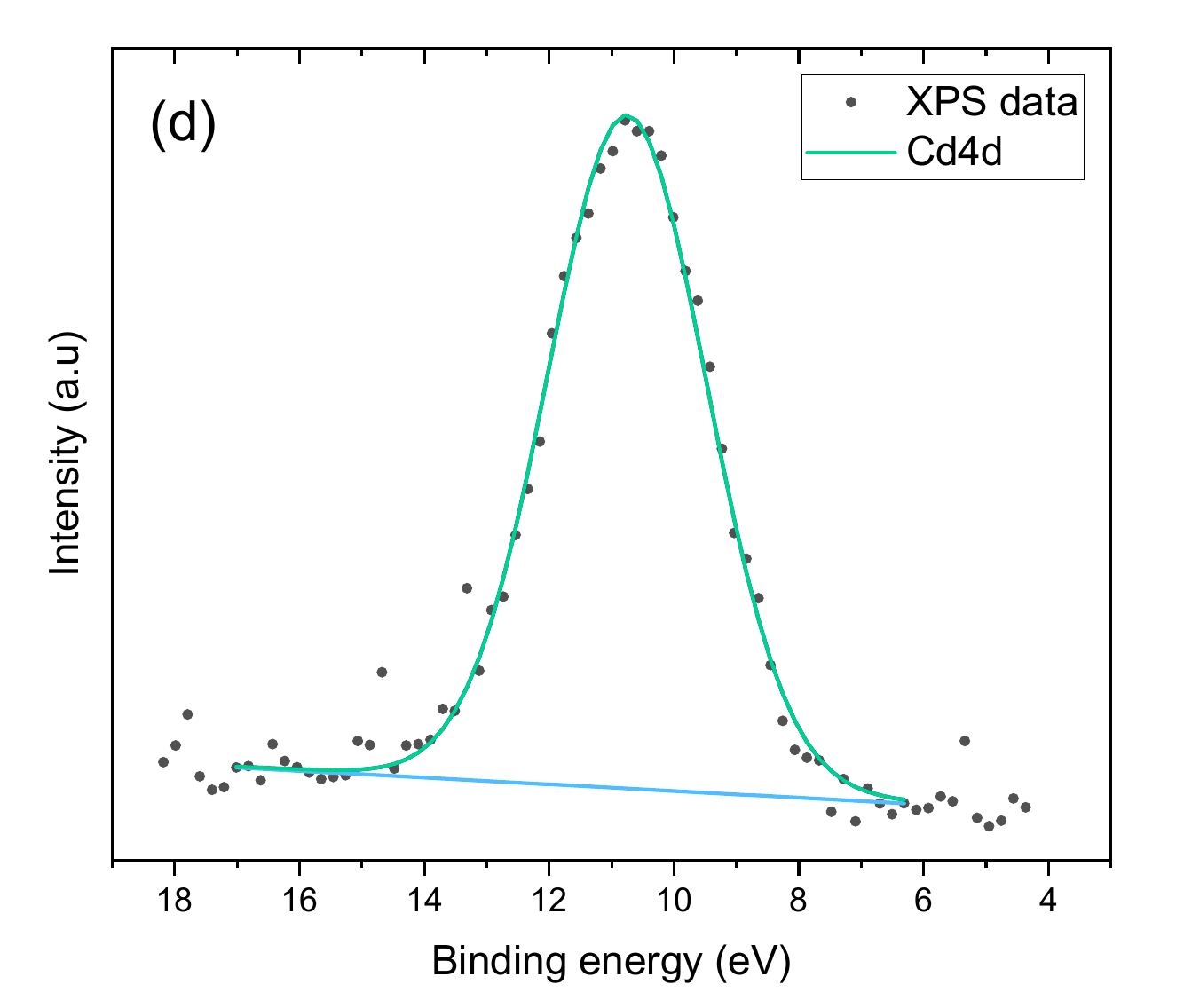} 
        \label{fig15d}
	\end{minipage} 
 \hspace{1cm} 
	\begin{minipage}[t]{6cm} 
		\centering 
		\includegraphics[scale=0.3]{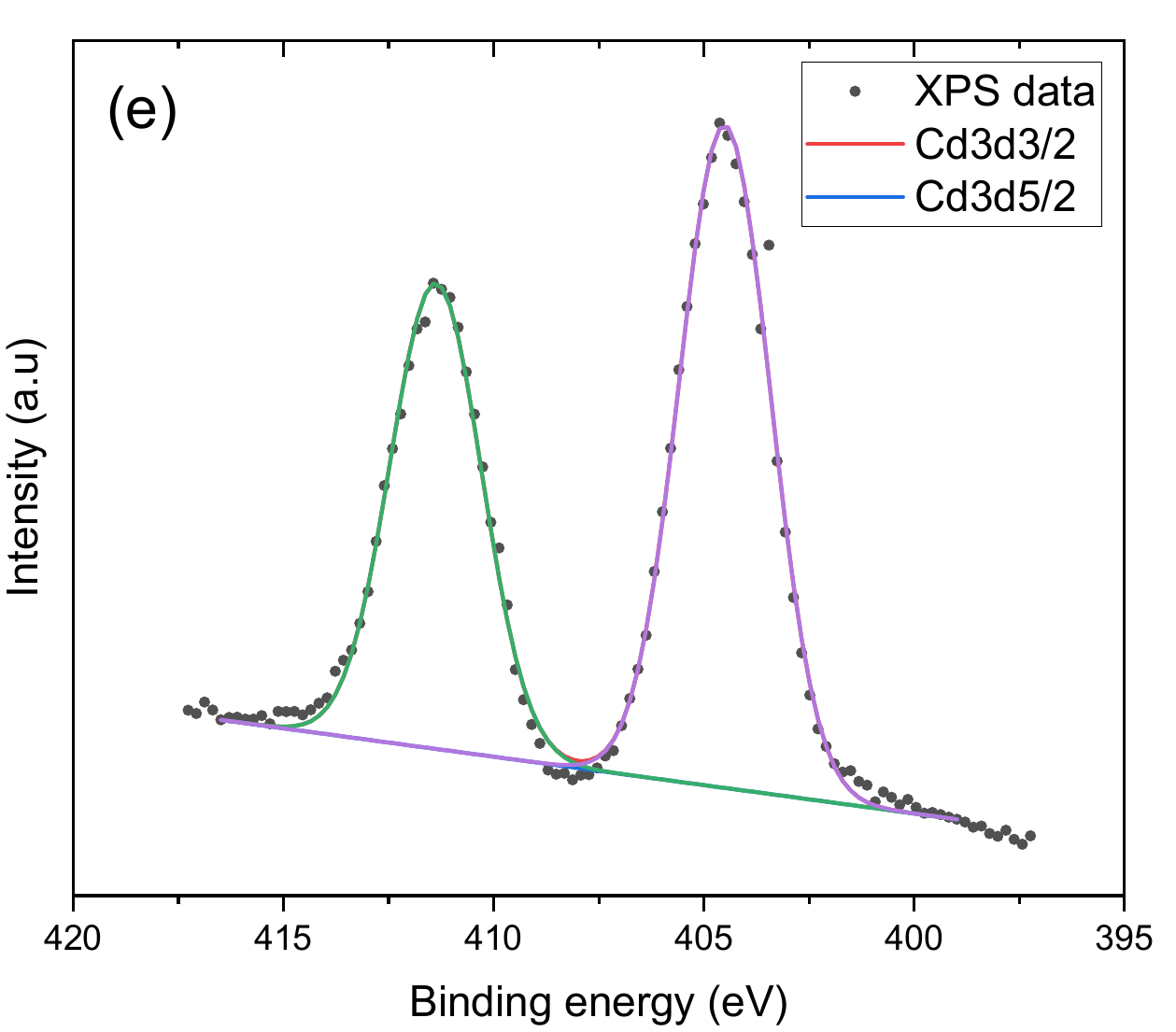} 
        \label{fig15e}
	\end{minipage} 
 \caption{(b),(c) XPS spectra of C1s and O1s for the calibration of raw specrum, (d) and (e) XPS spectra of Cd4d and Cd3d5/2 respectively}
 \end{figure}
 
XPS measurements were used to examine the surface chemical composition of the prepared target. For these measurements, the XPS setup provided by VSW Ltd., UK, was utilized. A base pressure of 5 $\times$ 10$^{-10}$ mbar was attained with the aid of a 1250 \textit{l}/s turbo molecular pump, complemented by a dry-scroll pump during the measurement process. The chamber and X-ray gun were connected to a Varian, Inc. ion pump through a Tee setup, where the X-ray gun was differentially evacuated. A monochromatic Mg K$_\alpha$ X-ray source with an energy of 1253.6 eV was utilized for the XPS analysis. The XPS unit comes with a twin-anode X-ray gun and a hemispherical e-analyzer with a radius of 150 mm. Electrons emitted from the target are extracted by a cylindrical electromagnetic lens and directed into the hemispherical analyzer's entrance slit. To capture the photoelectrons, a multichannel detection (MCD) device was used. Using a 20 eV pass energy, XPS spectra of Cd 4d, Cd 3d$_{3/2}$, Cd 3d$_{5/2}$, C 1s, and O 1s were obtained within different energy windows. The energy resolution of the XPS system was 1.88 eV, and no charge neutralizer was used throughout the measurements. The XPS spectra were analysed using the XPSpeak41 software, and the binding energy of the C 1s and O 1s peaks was utilised to calibrate the raw spectrum's energy \cite{xps-greczynski2020x}\cite{achary2014one}.

\subsection{X-ray Fluorescence (XRF)}

\begin{figure}[!h] 
\centering 
	\begin{minipage}[t]{6cm} 
		\includegraphics[scale=0.08]{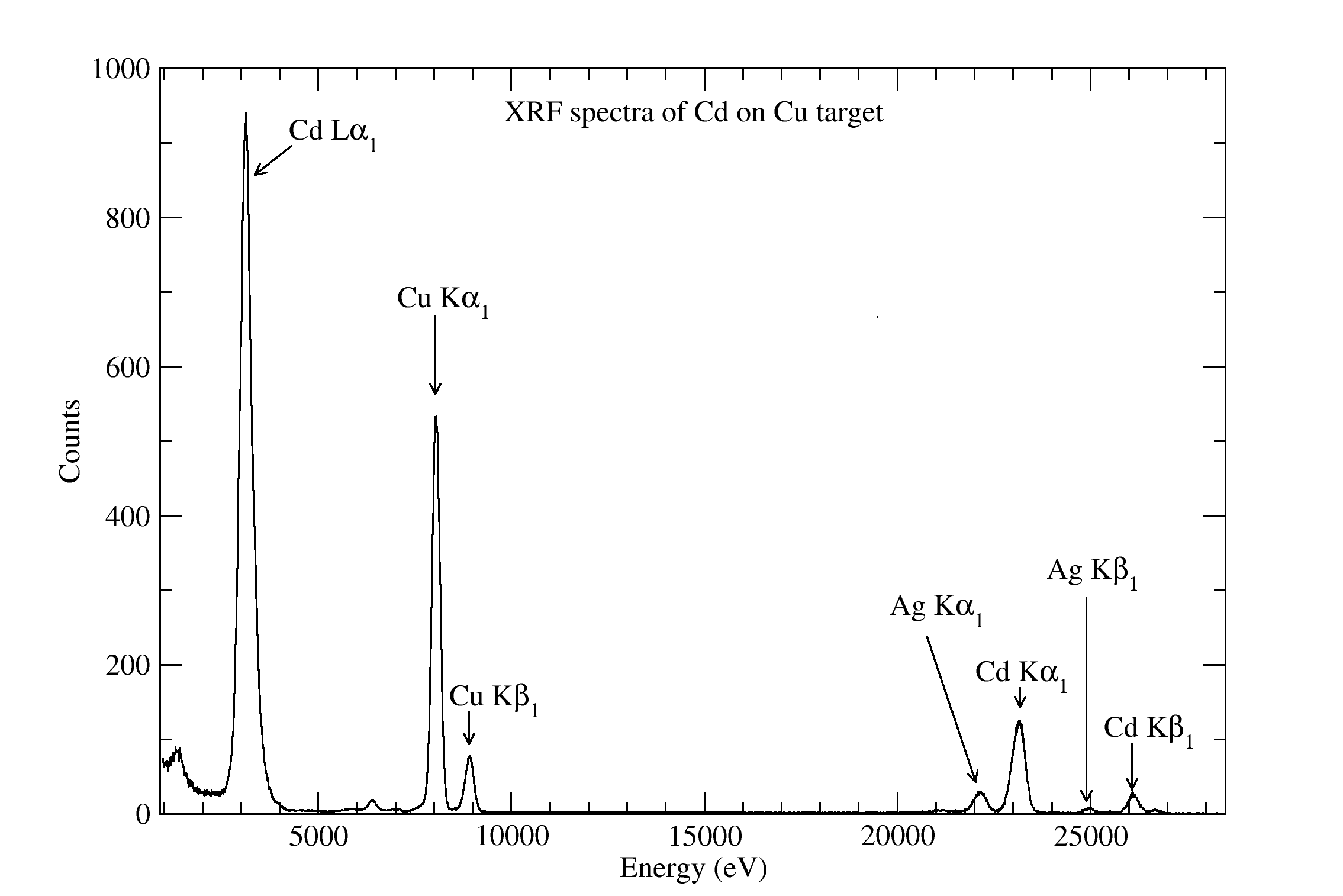} 
        \label{fig16a}
	\end{minipage} 
	\begin{minipage}[t]{6cm} 
		\includegraphics[scale=0.08]{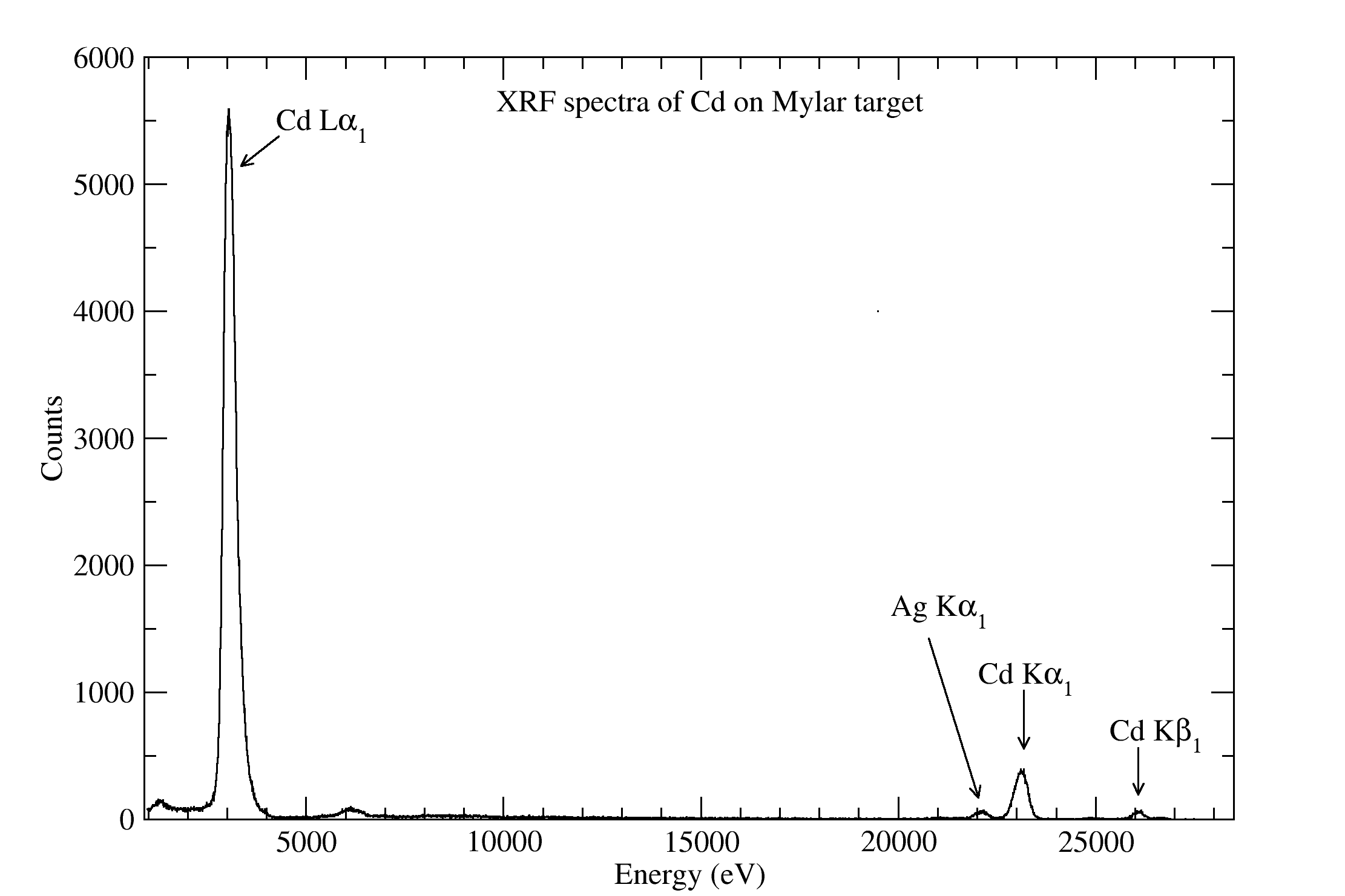}  
        \label{fig16b}
	\end{minipage} 
\caption{XRF spectra of `Cd on Cu' and `Cd on mylar' targets respectively. Ag K$\alpha$$_1$ and Ag K$\beta$$_1$ peaks are coming from machine parts}
 \end{figure}
 
The XRF (X-ray Fluorescence) technique was used for the qualitative elemental characterization of the Cd on Mylar and Cd on Cu targets. The experimental setup, built by Moxtek,USA, consists of a 4W MAGNUM X-ray source, a Si-pin detector, and a digital pulse processor. Figure 1 shows a schematic diagram. The X-ray tube is equipped with an Ag anode and a 0.25mm thick beryllium window. The Si-pin detector has a DuraBeryllium window 25 mm thick, with a detector thickness and active area of 625$\mu$m and 6mm$^2$, respectively. The MXDPP50 digital pulse processor, which includes a 4k channel MCA, a detector temperature controller, and a detector power supply, aided in data collecting. The SinerX software was used to control the digitizer, detector, and X-ray tube parameters \cite{gupta2017characterization}.

The presence of Cd in both targets is confirmed by the Cd L$\alpha_1$, Cd K$\alpha_1$, and Cd $\beta_1$ lines of Cd at 3100 eV, 22180 eV and 26102 ev respectively. In the Cd on Cu backing targets, the Cu K$\alpha_1$, and Cu K$\beta_1$ of Cu are attributed to the Cu backing of the target, and no impurity of heavy mass elements are present. The Ag lines are observed due to the presence of the Ag anode in the X-ray tube. The detector was calibrated using a Cu coin, and the experiment was conducted in an on-air condition.

\section{$^{108}$Cd(p,$\gamma$)$^{109}$In reaction with prepared targets}
\label{exp}
\begin{figure}[!h]
     \centering
    \includegraphics[scale=0.4]{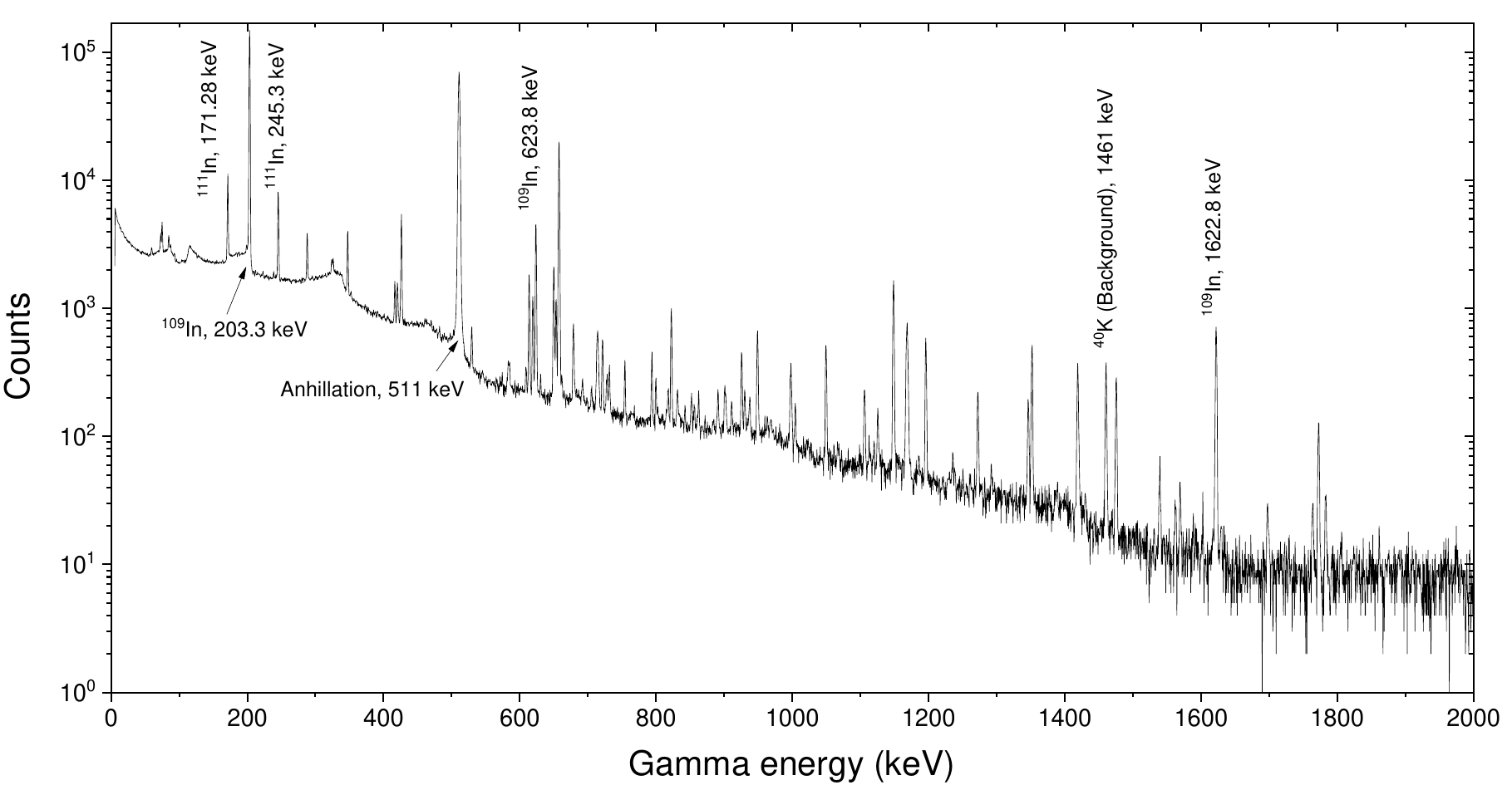}
    \caption{The offline $\gamma$-ray spectrum from Cadmium (66.3$\%$ enriched $^{108}$Cd) on Mylar target irradiated by 5.19 MeV of proton beam}
    \label{fig18}
\end{figure}

The cross-section of $^{108}$Cd(p,$\gamma$)$^{109}$In was determined through activation techniques. A 7 MeV proton beam was provided by the K130 cyclotron at Variable Energy Cyclotron Cetre (VECC), Kolkata, India and then degraded to 6.85 MeV to 2.27 MeV using 99.95\% pure Al foils. Each target setup was irradiated for a duration of 10-15 hours at a beam current of 150 enA. Following the irradiation period, the targets were allowed to cool to minimize unnecessary gamma peaks. Subsequently, the targets were placed in front of an HPGe detector at the Analytical Chemistry Division, BARC at VECC, Kolkata, India for gamma counting. Throughout the irradiation, a constant flow of chilled water was maintained for target cooling. After irradiation, no damage was observed on the targets.

Figure \ref{fig18} illustrates the spectrum of offline gamma rays measured from target number Cd3 (Cadmium on Mylar) irradiated by a 5.19 MeV proton beam for 6 hours and 7 minutes. The $\gamma$-rays were recorded for 1003 seconds after a 2-hour cooling time. The target comprises 29.1$\%$ of $^{110}$Cd. In the reaction $^{110}$Cd(p,$\gamma$)$^{111}$In, $^{111}$In produces $\gamma$-rays at 171.28 keV and 245.35 keV with relative intensities of 90.7$\%$ and 94.1$\%$, respectively. Additionally, from the reaction $^{108}$Cd(p,$\gamma$)$^{109}$In, $^{109}$In generated predominantly emits $\gamma$-rays with relative intensities greater than 0.2$\%$, with the most prominent ones being at 203.3 keV and 623.8 keV, having relative intensities of 74.2$\%$ and 5.64$\%$ followed by $\beta$$^+$ decay with a half-life of 4.159 hours \cite{1-gyurky2006106}. Other gamma rays originate from natural radioactive isotopes.

\section{Summary and conclusion}
\label{}
This paper presents the fabrication method of $^{108}$Cd on Mylar and $^{108}$Cd on copper backing. The optimization process involved adjusting parameters such as the distance between the crucible and substrate holder, the dimensions of the crucible opening, the e-beam current, and the evaporation time. Ultimately, 38.7 mg of 66.3$\%$ enriched $^{108}$Cd was utilized with a beam current ranging from 3 to 7 mA and a deposition time of 30 minutes. The separation between the crucible and substrate holder was set at 5 cm. The thickness of the cadmium targets was determined through $\alpha$-energy loss measurements and validated by RBS measurements. We utilized eight different Cadmium targets with thicknesses ranging from 0.33 $\mu$m (surface density, 288 $\mu$g/cm$^2$) to 0.76 $\mu$m (656 $\mu$g/cm$^2$) on a Mylar backing during the $^{108}$Cd(p,$\gamma$)$^{109}$In reaction measurement.

The non-uniformity of the target thickness was observed to be 10$\%$. Adjustments to the thickness of the targets can be made by using an appropriate amount of material during the evaporation process. Alternatively, other processes, such as electrodeposition, can be used for the fabrication of cadmium targets \cite{scholz2016constraints}\cite{bar2022preparation}\cite{biswas2023fabrication}.

The absence of impurities was confirmed through RBS measurements. XPS and XRF measurements provided insights into the quality of the prepared targets, evaluating both surface morphology and bulk composition. The targets exhibited no significant elemental impurities, apart from the presence of carbon, hydrogen and oxygen.

\section{Acknowledgements}
\label{}

The authors express their gratitude to Prof. Supratic Chakraborty of the Surface Physics and Material Science division (SPMS), Saha Institute of Nuclear Physics (SINP), Kolkata, India, for their assistance in preparing the target using the Telemark multipocket e-beam facility. Special thanks to Mr. Abhilash S.R. from the Inter-University Accelerator Centre, New Delhi, India, and Gy. Gyurky from ATOMKI, Hungary, for insightful discussions regarding the fabrication of Cadmium. The authors acknowledge Mr. Debraj Dey and Mr. Gautam Sarkar from SPMS division, SINP, for their assistance with XPS measurements. The authors thank to Mr. Arkabrata Gupta from the Indian Institute of Engineering Science and Technology, Shibpur, India, for helping with the XRF measurements. The SINP workshop group is appreciated for their contributions to the setup modifications. Sukhendu Saha expresses thanks to the Council of Scientific \& Industrial Research (CSIR), India, for financial support through Senior Research Fellowship (File no: 09/489(0119)/2019-EMR-I).





\bibliographystyle{elsarticle-num-names}
\bibliography{target_preparation}

\end{document}